\RequirePackage[hyphens]{url}
\documentclass[a4paper,fleqn]{cas-dc}
\usepackage{cas-dfrws} 
\renewcommand{\dfrwsconference}{Digital Forensics Research Conference (DFRWS EU), 2024}
\pdfoutput=1

\usepackage{stix}
\usepackage{paralist}
\usepackage{graphicx}
\usepackage{amsmath}
\usepackage{amsthm}
\usepackage{lipsum}

\usepackage[dvipsnames]{xcolor}
\usepackage{mathtools}
\usepackage[colorlinks]{hyperref}
\hypersetup{
    linkcolor=CornflowerBlue,
    citecolor=CornflowerBlue,
    urlcolor=cyan
}
\usepackage{cleveref}
\usepackage{array,multirow}
\usepackage{caption, subcaption}
\usepackage[binary-units]{siunitx}
\usepackage{enumitem}
\usepackage{listings}
\lstdefinestyle{basic}
{
    frame=tb,
    rulecolor=\color{lightgray},
    backgroundcolor=\color{black!1!},
    basicstyle=\footnotesize\color{black}\ttfamily,
    commentstyle=\color{commentsColor}\textit,
    breaklines=true,
    aboveskip=0.2cm,
    belowskip=0.0cm,
    escapeinside={\%*}{*)}
}

\colorlet{punct}{red!60!black}
\definecolor{delim}{RGB}{20,105,176}
\colorlet{numb}{magenta!60!black}

\lstdefinelanguage{json}{
    basicstyle=\normalfont\ttfamily,
    showstringspaces=false,
    breaklines=true,
    frame=lines,
    backgroundcolor=\color{black!3},
    literate=
     *{0}{{{\color{numb}0}}}{1}
      {1}{{{\color{numb}1}}}{1}
      {2}{{{\color{numb}2}}}{1}
      {3}{{{\color{numb}3}}}{1}
      {4}{{{\color{numb}4}}}{1}
      {5}{{{\color{numb}5}}}{1}
      {6}{{{\color{numb}6}}}{1}
      {7}{{{\color{numb}7}}}{1}
      {8}{{{\color{numb}8}}}{1}
      {9}{{{\color{numb}9}}}{1}
      {:}{{{\color{punct}{:}}}}{1}
      {,}{{{\color{punct}{,}}}}{1}
      {\{}{{{\color{delim}{\{}}}}{1}
      {\}}{{{\color{delim}{\}}}}}{1}
      {[}{{{\color{delim}{[}}}}{1}
      {]}{{{\color{delim}{]}}}}{1},
}

\usepackage{tikz}
\usetikzlibrary{shapes.multipart}
\usetikzlibrary{arrows,calc,shapes,decorations.pathreplacing,matrix,calligraphy}
\usetikzlibrary{decorations.markings}
\usetikzlibrary{backgrounds,positioning,automata,arrows.meta}
\usetikzlibrary{tikzmark}

\usepackage{pgfplots}
\pgfplotsset{compat=newest}
\usepackage[framemethod=TikZ]{mdframed}
\mdfdefinestyle{TikzFrame}{%
    roundcorner=2pt,
    innertopmargin=4pt,
    innerbottommargin=4pt,
    innerrightmargin=0pt,
    innerleftmargin=4pt,
    leftmargin = 4pt,
    rightmargin = 4pt
}

\usepackage[authoryear]{natbib}
\begin{document}

\title[mode=title]{Nyon Unchained: Forensic Analysis of Bosch's eBike Board Computers}
\shorttitle{Nyon Unchained: Forensic Analysis of Bosch's eBike Board Computers}

\shortauthors{Stachak et~al.}

\author[1]{Marcel Stachak}
\ead{marcel.stachak@fau.de}
\credit{Conceptualization, Methodology, Investigation, Writing - Review and Editing}

\author[1]{Julian Geus}[orcid=0009-0001-8270-1964]
\cormark[1]
\ead{julian.geus@fau.de}
\credit{Conceptualization, Methodology, Validation, Investigation, Writing - Original Draft, Writing - Review and Editing, Supervision}

\author[1]{Gaston Pugliese}
\ead{gaston.pugliese@fau.de}
\credit{Investigation, Resources, Writing - Review and Editing, Supervision}

\author[1]{Felix Freiling}[orcid=0000-0002-8279-8401]
\ead{felix.freiling@fau.de}
\credit{Writing - Review and Editing, Supervision}

\address[1]{Department of Computer Science,
  Friedrich-Alexander-Universit\"at Erlangen-N\"urnberg (FAU),
  Erlangen, Germany}

\cortext[1]{Corresponding authors.}

\begin{abstract}
    Modern eBike on-board computers are basically small PCs that not only offer motor control,
    navigation, and performance monitoring, but also store lots of sensitive user data.
    The Bosch Nyon series of board computers are cutting-edge devices from one of the market
    leaders in the eBike business, which is why they are especially interesting for forensics.
    Therefore, we conducted an in-depth forensic analysis of the two available Nyon models released in 2014 and 2021. 
    On a first-generation Nyon device, Telnet access could be established
    by abusing a design flaw in the update procedure, 
    which allowed the acquisition of relevant data without
    risking damage to the hardware. Besides the user's personal information, 
    the data analysis revealed databases containing user activities, 
    including timestamps and GPS coordinates.
    Furthermore, it was possible to forge the data on the device 
    and transfer it to Bosch's servers to be persisted across their online service and smartphone app.
    On a current second-generation Nyon device, no
    software-based access could be obtained. For this reason, more intrusive hardware-based options
    were considered, and the data could be extracted via chip-off eventually. 
    Despite encryption, the user data could be accessed and evaluated.
    Besides location and user information, the newer model holds
    even more forensically relevant data, such as nearby Bluetooth devices.
\end{abstract}


\maketitle


\section{Introduction}
\label{introduction}

Despite constant---and partly spectacular
\citep{DBLP:journals/compsec/Garfinkel13}---advances
\citep{DBLP:books/daglib/0027500,DBLP:journals/dt/FreilingGLMP18} 
in the area of digital forensic science, the analysis of concrete
digital devices remains a cumbersome undertaking
\citep{DBLP:journals/di/Garfinkel10}. While much standardization has
occurred, especially with respect to interfaces to devices (e.g.,
JTAG, USB) and storage (e.g., SATA, MMC), the variation of devices
themselves has increased. A specific driving force behind this
development is the proliferation of \emph{special-purpose mobile
  devices}, i.e., devices specially built to fulfil a certain task or
purpose. The forensic analysis of such devices is the focus of this
paper.

In their \emph{non}-mobile form, such special-purpose devices usually
run under the heading of \emph{embedded systems} and have appeared in
the form of point-of-sale terminals, cash machines, or information
displays. In contrast to general-purpose \emph{mobile} devices like
smartphones, special-purpose mobile devices are \emph{mobile} embedded
systems with a specific purpose. Examples are navigation systems,
fitness trackers, or robot vacuum cleaners. 
The \emph{Internet of Things} (IoT), as a buzzword, 
is often used to characterize special-purpose embedded systems.

An interesting type of special-purpose mobile devices has been on the
market for almost 10 years: Bike computers are the information hub
for electric bikes, also known as ``pedelecs''.
According to~\citet{ebike_stats}, pedelecs have a rapid increase in sales,
which is caused by the broad target market
that covers all age groups, and since they are often a cheaper
and a more environmental friendly alternative to cars.
Their ever-growing numbers and increasing relevancy in modern day
transportation further supports the importance of considering their
accompanying bike computers as sources of forensically relevant evidence.
One of the market leaders of digital technology in
Germany is the company Bosch. Bosch nowadays speaks about ``connected
biking''. Especially their most feature-rich series of bike computers,
called \emph{Nyon}, might store lots of user-specific data and is therefore
the focus of our analysis.

\subsection{Related Work}
\label{sec:related:work}

Research in the field of non-classical PC forensics is necessarily
based on specific case studies and many such studies have been
performed. One of the earliest and most well-known studies, with more focus on
reverse engineering than on forensic analysis, was performed when
\citet{xbox:2003} ``hacked the Xbox''. On the mobile side of gaming
consoles and with focus on forensic analysis, \citet{barr2021dead}
analyzed the portable video gaming console Nintendo Switch. The
authors used an exploit to gain access to the device and acquire a
memory dump. They meticulously analyzed the data and were able to
identify numerous forensically interesting traces.

In the field of IoT forensics, \citet{youn2021forensic} analyzed
Amazon's AI speaker Echo Show.  The authors extracted the data via 
chip-off and also took smartphone data from the companion app into
account. In contrast, \citet{villarreal2022nondestructive} developed
a non-destructive method to acquire data from the flash memory chip of
Amazon's Echo Dot devices.

Cars and their built-in infotainment systems are an interesting case
since those parts of the system that coordinate the brakes and
steering have a special purpose while the generic parts of the
automotive computing system that are responsible for entertainment
typically allow general-purpose computations. Automotive digital
forensics \citep{DBLP:journals/tiv/StrandbergNO23} has been of special
interest for many years. While earlier work by
\citet{DBLP:conf/safecomp/HoppeKKD12} focused on extracting route
information from the car itself, \citet{ebbers2021grand} attempted to
extract the same data from the companion apps of car manufacturers.

Due to the heterogeneity of these devices, 
it is hard to come up with practical yet generic analysis
guidelines. \citet{DBLP:journals/di/GomezMGM21} proposed a generic
investigation methodology for IoT devices, while
\citet{buquerin2021generalized} evaluated digital forensic processes
in the automotive domain and developed a general process for
automotive digital forensic investigations.  Especially because of the
widespread usage of encryption, as well as other security measures,
\citet{fukami2021new} proposed a model for data acquisition from
mobile devices with particular emphasis on smartphones.  
In special-purpose mobile devices, the security might not be as advanced,
therefore, this model only partially applies. Still, specific-purpose
mobile devices are usually feature rich, often run a full Linux or
Android operating system, but do not allow running custom software
like apps. A generic data acquisition strategy for more
feature-restricted devices is still an open research question.

\subsection{Contributions}

Generally, we are not aware of any work that has 
analyzed eBike computers for their forensic or security aspects. 
Therefore, the contributions of this paper are as follows: 

\begin{compactitem}

\item To the best of our knowledge, we are the first to forensically
  analyze the first- and second-generation versions of the Bosch Nyon
  eBike board computer.

\item We identified a design flaw in the update process of the Nyon~2014 that
    enables data acquisition without the need for intrusive
	hardware-based options.

\item We developed a data acquisition methodology for special-purpose
	mobile devices, including an assessment of the forensic requirements.

\item We conducted an in-depth data analysis, highlighting the forensic value of
    special-purpose mobile devices.

\end{compactitem}

\subsection{Outline}

In \Cref{sec:devices}, we provide background information on Bosch's Nyon devices, 
including hardware and software details. 
In \Cref{sec:method}, we address challenges for data acquisition on mobile devices,
and propose a methodology adapted to the Nyon devices. 
Based on this methodology, the analysis results for the first-generation Nyon device 
are reported in \Cref{sec:nyon14}, 
and for the second-generation Nyon device in \Cref{sec:nyon21}. 
After discussing our findings in \Cref{sec:implications}, 
we conclude the paper in \Cref{sec:conclusion}.


\begin{figure}[tb]
    \centering
    \begin{subfigure}[b]{0.3\textwidth}
	    \includegraphics[width=\textwidth]{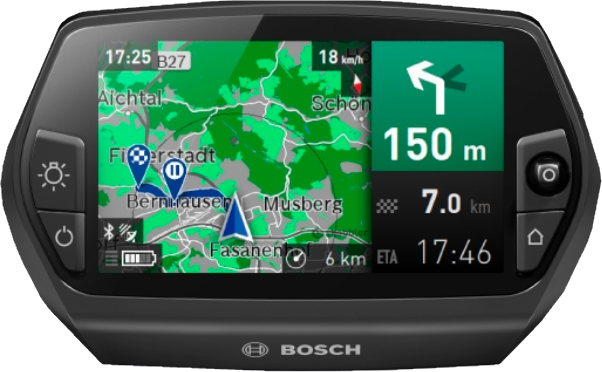}
	    \caption{Nyon 2014}
	    \label{fig:nyon14}
    \end{subfigure}
    \hfill
    \begin{subfigure}[b]{0.15\textwidth}
	    \includegraphics[width=\textwidth]{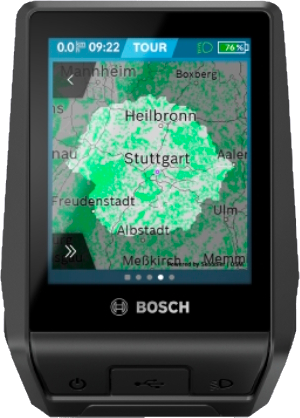}
	    \caption{Nyon 2021}
	    \label{fig:nyon21}
    \end{subfigure}
    \caption{Bosch Nyon computers (\textcopyright{} \emph{Robert Bosch GmbH}).}
    \label{fig:nyons}
\end{figure}

\section{Bosch Nyon Computers}
\label{sec:devices}

Nyon devices are the premium class of portable bike computers by Bosch 
for electronic bicycles with supported ``eBike drives''. 
Since 2014, Bosch released two models of the Nyon computer 
which are shown in \autoref{fig:nyons}. 
Both of them are connected to the electronics of the eBike 
through a docking station located on the bike's handlebar.

\subsection{Bosch Nyon 2014}
\label{sec:nyon14basics}

The first generation of the Bosch Nyon computer was released in 2014 
and offered with \SI{1}{\giga\byte} (``BUI270'') 
and \SI{8}{\giga\byte} (``BUI275'') of internal storage. 
The graphical user interface on the 4.3-inch non-touch screen (480x270 pixel)
can be controlled via an analog joystick and three buttons. 

\begin{figure}[tb]
    \centering
    \includegraphics[width=0.4\textwidth]{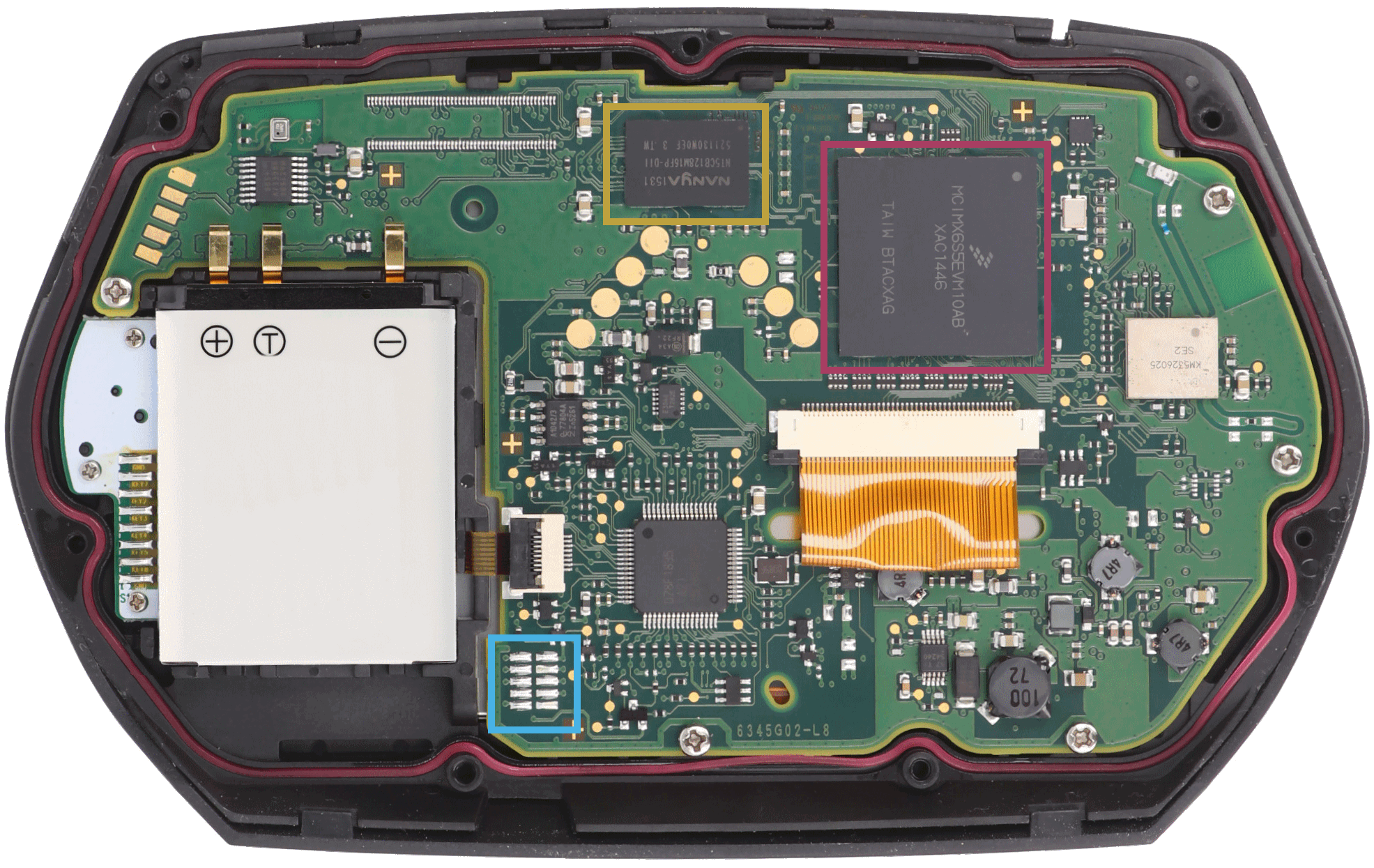}
    \caption{Main logic board (MLB) of the Nyon~2014 highlighting the  
    \textcolor{red!70!black}{CPU}, \textcolor{yellow!70!black}{DRAM}, and a 
    \textcolor{blue!50!green}{possible debug interface}.}
    \label{fig:nyon1board}
\end{figure}

\autoref{fig:nyon1board} shows the main logic board (MLB) 
of the ``BUI275'' version of the Nyon~2014 after removing the rear casing, 
and reveals the following components:

\begin{compactitem}
\item i.MX 6Solo CPU (MCIMX6S5EVM10AB) by \cite{nxp:MCIMX6S5EVM10AB} with 32-bit ARM Cortex-A9,
\item two DDR3(L) SDRAM chips (NT5CB128M16FP-DII) by \cite{nanya:NT5CB128M16FP_DII} with \SI{128}{\mega\byte} each, 
\item eMMC with \SI{8}{\giga\byte} (FBGA code ``JWB18'') by \cite{micron:JWB18}.
\end{compactitem}

The open-source software utilized in ``eBike Systems'' products 
can be determined on Bosch's license website \citep{bosch_oss}. 
Thereby, it is revealed that the Nyon~2014, for instance, 
is based on Linux kernel 3.0.35, uses U-Boot as bootloader, 
and incorporates parts of the Android Open Source Project (AOSP). 
During our analysis of the Nyon~2014 in \Cref{sec:nyon14}, 
we further identified the Linux distribution to be ``Poky Linux'',
a reference distribution of the \cite{yocto:poky} developed for embedded systems.

The micro USB port of the Nyon~2014 can be utilized 
to charge the device or to connect it to a PC running Bosch's
``DiagnosticTool'', the debugging and diagnosis software 
for eBike manufacturers or retailers. 
An Internet connection can be established with the built-in Wi-Fi module 
to synchronize data to Bosch's cloud or to download map data.
Via Bluetooth, the Nyon can be paired with a smartphone running the 
``Bosch eBike Connect'' app to obtain mobile Internet access. 
Bluetooth can also be utilized to connect supported peripherals  
which are, however, limited due to the missing support for Bluetooth Low Energy (BLE).

\subsection{Bosch Nyon 2021}
\label{sec:nyon21basics}

The second-generation and latest Nyon device 
is known as ``BUI350'' and was released in 2021. 
Its more compact design and portrait-format 3.2-inch touch screen (370x454 pixel)
distinguishes the Nyon~2021 computer 
from its first-generation predecessor (cf. \autoref{fig:nyons}). 

\begin{figure}[tb]
    \centering
    \includegraphics[width=0.2\textwidth]{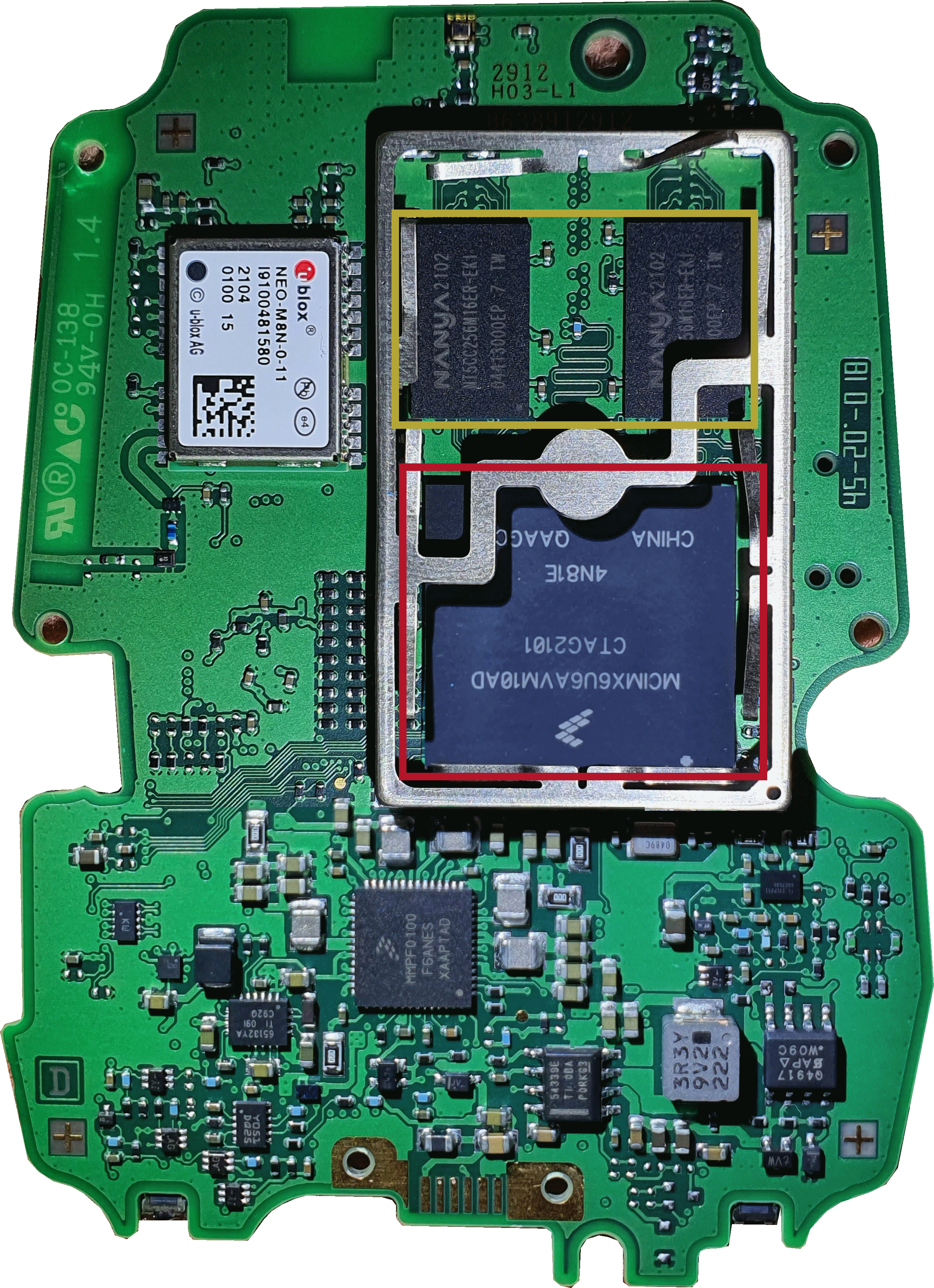}
    \hspace{0.4cm}
    \includegraphics[width=0.2\textwidth]{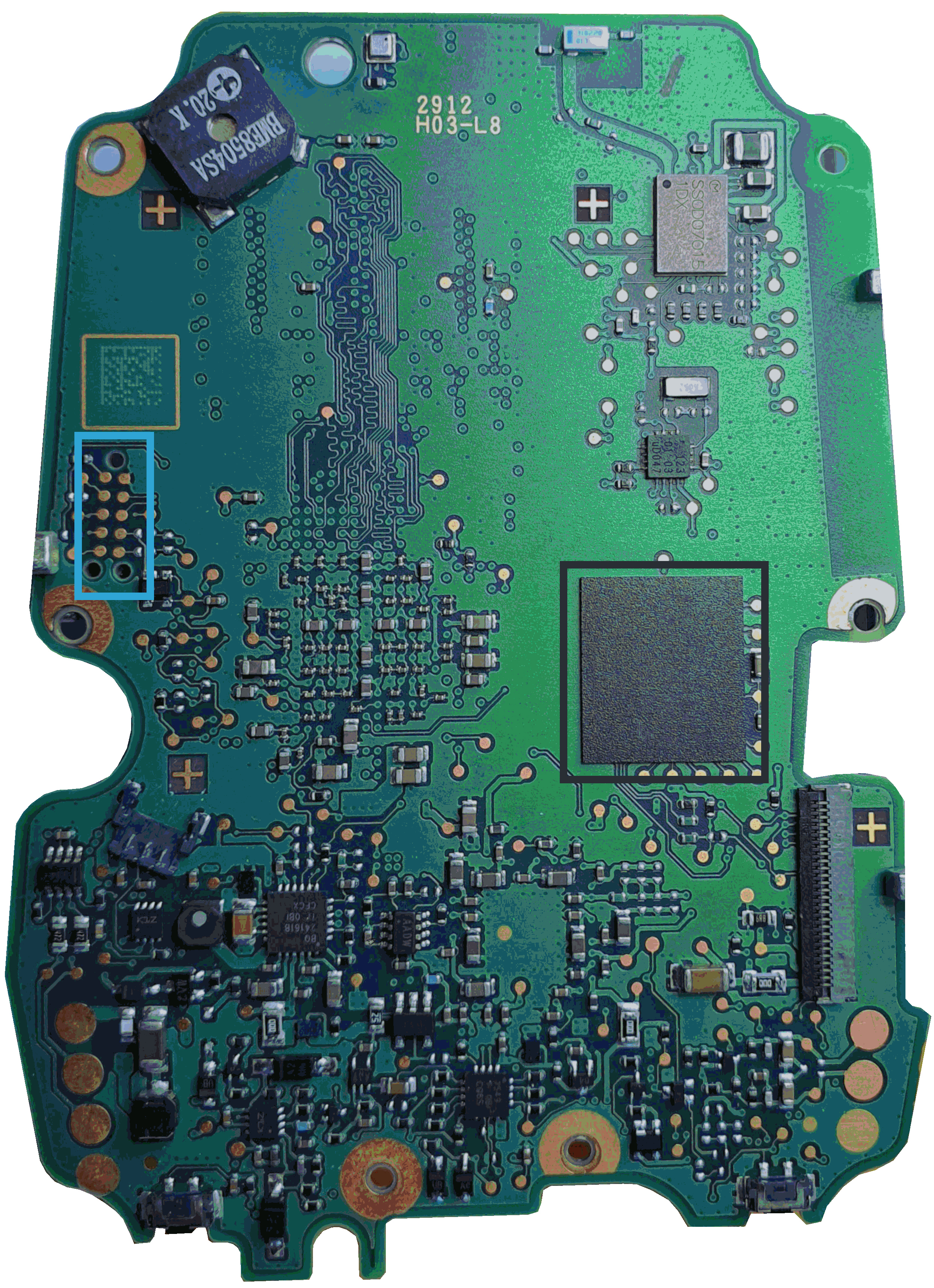}
    \caption{Main logic board (MLB) of the Nyon~2021 highlighting the 
    \textcolor{red!70!black}{CPU}, 
    \textcolor{yellow!70!black}{DRAM}, 
    a \textcolor{blue!50!green}{possible debug interface}, and \textcolor{black!70}{eMMC}.}
    \label{fig:nyon2board}
\end{figure}

\autoref{fig:nyon2board} shows the front and back side of the MLB, 
and highlights the following identified components:

\begin{compactitem}
\item i.MX 6Solo CPU (MCIMX6U8DVM10AD) by \cite{nxp:MCIMX6S5EVM10AB} w/ 32-bit ARM Cortex-A9,
\item two DDR3(L) SDRAM chips (NT5CC256N16ER-EKI) by \cite{nanya:NT5CC256N16ER_EKI} w/ \SI{256}{\mega\byte} each,
\item eMMC with \SI{8}{\giga\byte} (THGBMJG6C1LBAIL) by \cite{kioxia:THGBMJG6C1LBAIL}. 
\end{compactitem}

Bosch's license website~\citep{bosch_oss} discloses 
the use of Linux kernel 4.19.44, 
and our analysis in \Cref{sec:nyon21} again confirmed 
the usage of the Poky Linux distribution (cf. \Cref{sec:nyon14basics}).
Contrary to its predecessor, however, 
the Nyon~2021 has more advanced security mechanisms activated:   
The Linux kernel feature ``dm-verity'' checks the integrity of selected partitions on boot 
to ensure an uncompromised OS. ``AppArmor'' allows fine-grained privilege settings, 
restricts access to certain files or hardware devices for applications, 
and user data is encrypted using ``Linux Unified Key Setup'' (LUKS) with ``cryptsetup''. 
These security measures are similar to those of modern smartphones.

Connectivity-wise, the Nyon~2021 features a micro-AB USB port, 
Wi-Fi (2.4 GHz) with the now outdated 802.11b/g/n standard, 
and Bluetooth for the communication with the companion app. 
Due to its BLE support, the spectrum of pairable peripheral devices is broader compared to the first Nyon.

\subsection{Smartphone App and Web Service}
\label{sec:nyon:appweb}

\begin{figure}[htb]
    \centering
    \begin{subfigure}[b]{0.155\textwidth}
	    \includegraphics[width=\textwidth]{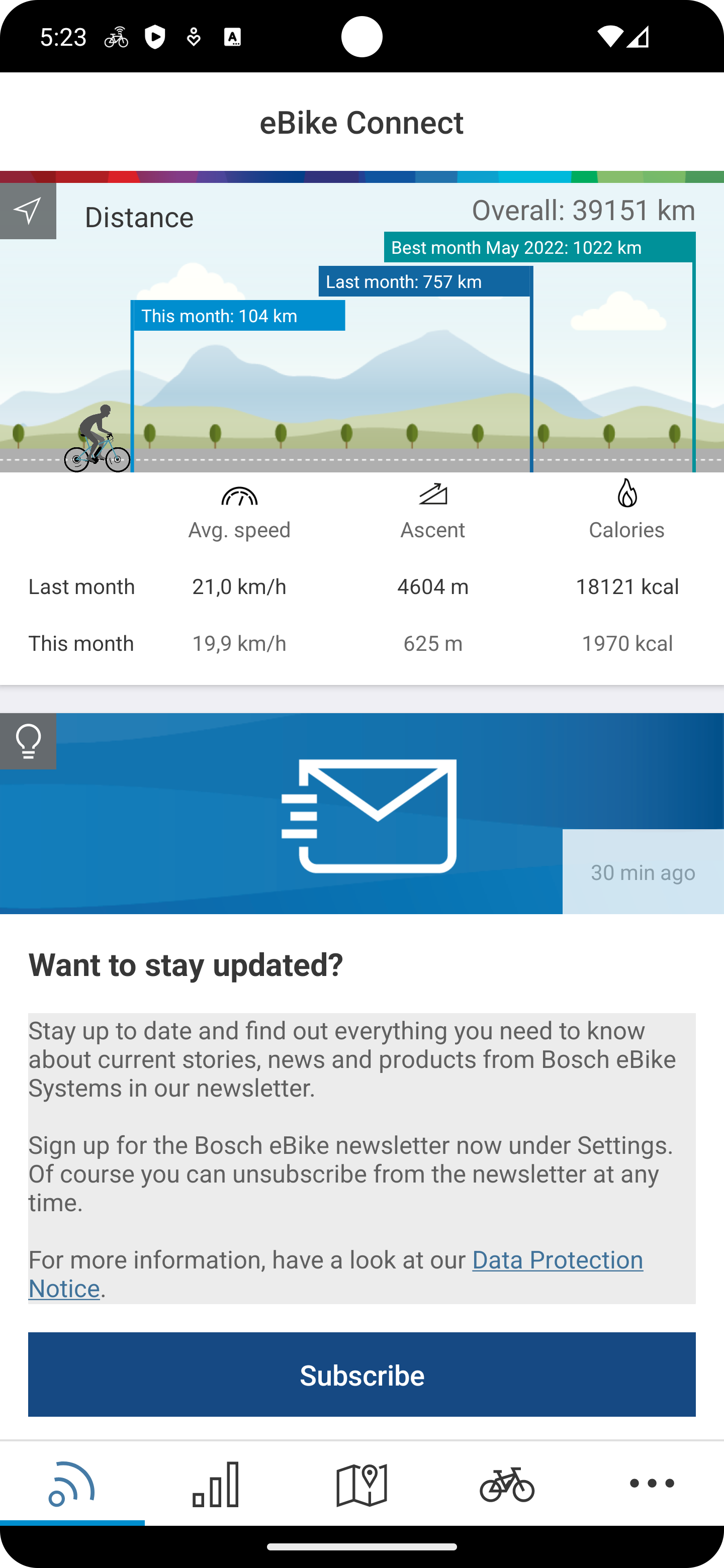}
	    \caption{Monthly statistics}
	    \label{fig:app1}
    \end{subfigure}
    \hfill
    \begin{subfigure}[b]{0.155\textwidth}
	    \includegraphics[width=\textwidth]{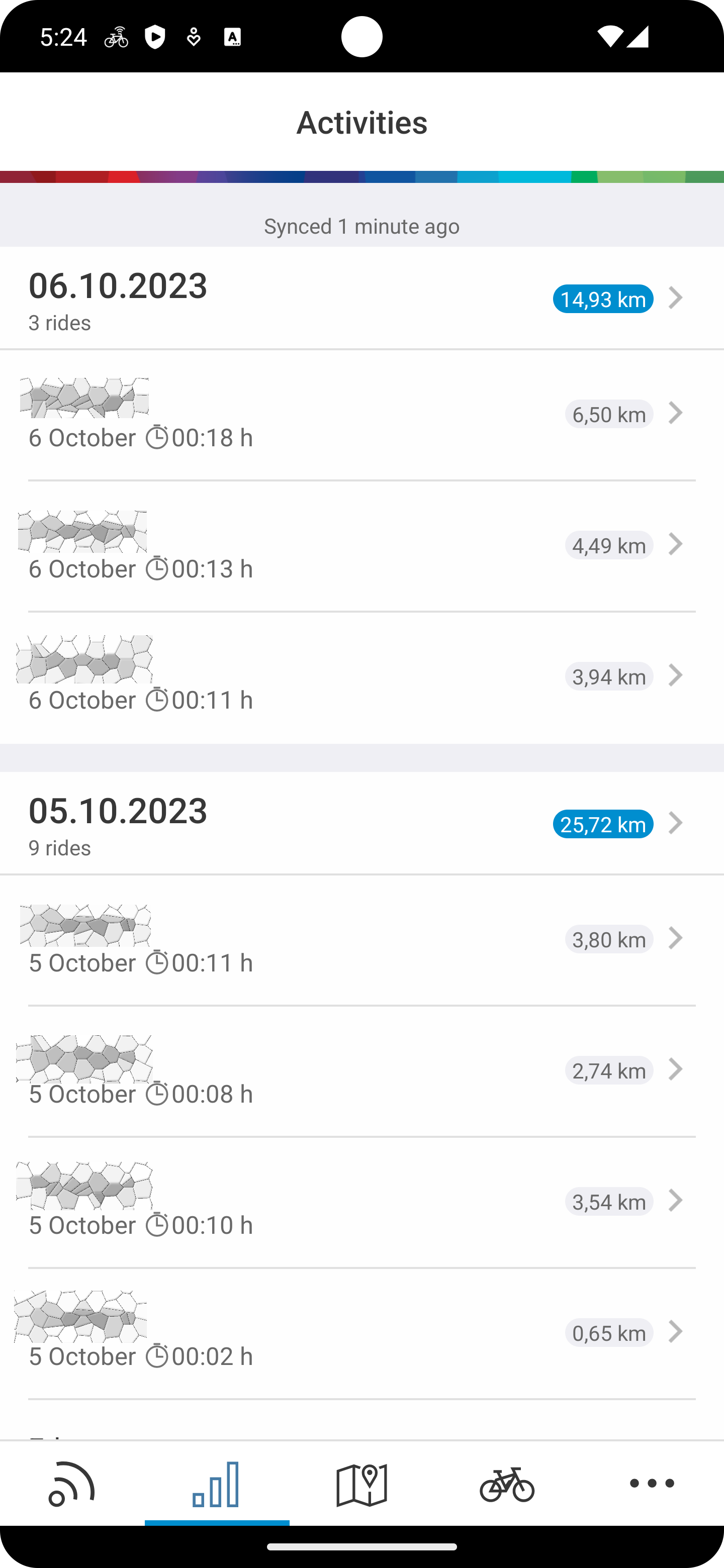}
	    \caption{Recent trips}
	    \label{fig:app2}
    \end{subfigure}
    \hfill
    \begin{subfigure}[b]{0.155\textwidth}
	    \includegraphics[width=\textwidth]{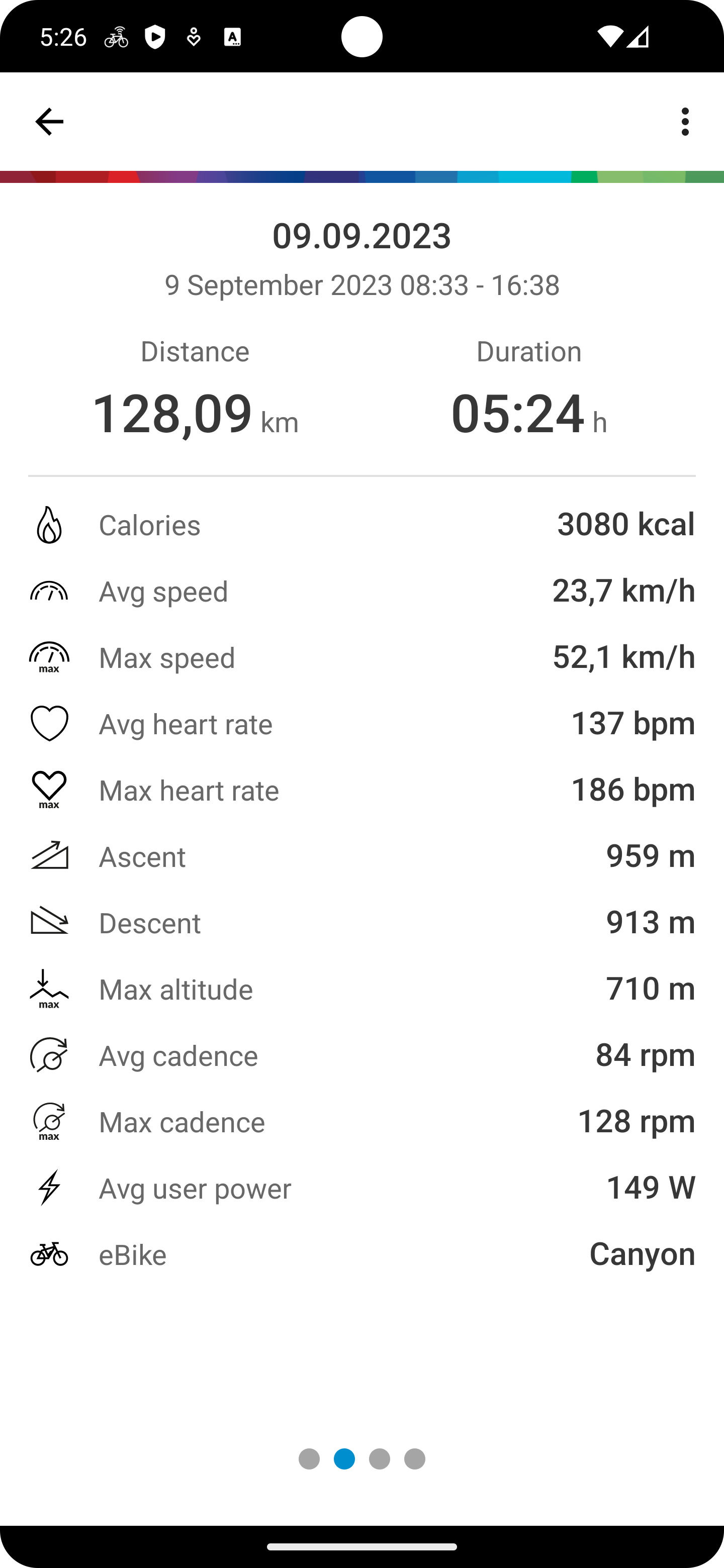}
	    \caption{Trip details}
	    \label{fig:app3}
    \end{subfigure}
    \caption{Screenshots of the eBike Connect app.}
    \label{fig:app}
\end{figure}

Nyon computers are accompanied by the smartphone app ``Bosch eBike Connect'' 
which is available for Android and iOS. 
As shown in \autoref{fig:app}, the app focuses on essential aspects 
related to users' cycling activities. 
While the first tab contains a monthly statistic on 
distances, average speeds, ascents, and calories (\autoref{fig:app1}),
the second tab provides an overview of recent trips on a daily basis (\autoref{fig:app2}), 
including a detail view for each of them which reveals additional information 
on speed, heart rate, altitude, or cadence (\autoref{fig:app3}). 
Moreover, users can plan trips and transfer them to the Nyon, 
as well as synchronize data with Bosch's cloud.

Another service by Bosch available for registered users is the 
``eBike Connect'' website~\citep{bosch_connect}. 
Although the functional scope of the website and the app are almost identical, 
the former additionally enables the export of trip data as GPX files, 
including all coordinates. 

Both the companion app and the web service require 
an ``eBike Connect'' online account which can be registered for free 
by providing country/language, first name, last name, email address, and a password.


\section{Data Acquisition Methodology}
\label{sec:method}

In classical digital forensics, the established paradigm of acquiring data
is to perform a bitwise copy of a hard drive. This approach is easy to
execute, ensures high data integrity, especially when the hard drive
is mounted read-only, can be easily verified and repeated, and provides
us with a comprehensive data set.  Since we are dealing with a
special-purpose mobile device, a lot of the same problems encountered
in mobile forensics also apply. Those challenges are described
by~\citet{fukami2021new}, which are mainly 
the widespread use of encryption and inaccessible key material, 
rendering acquisition by direct memory access hard or sometimes even impossible.  
This fact necessitates the use of imperfect acquisition methods, like the
device's backup mechanisms, as described by~\citet{geus23backup}.
However, the Nyon devices might not be as advanced as modern
smartphones when it comes to security, and therefore, we could face
only some of the same challenges.

Before we define a general acquisition methodology, we first investigate the main priorities
that have to be taken into account for forensic data acquisition, as well as the challenges with
those in mobile forensics:
\begin{itemize}
    \item \textbf{Integrity of Data}: One of the key points is to
	ensure the integrity of the data during the process. Therefore, the acquisition method
	has to be chosen carefully. However, in the mobile forensics sector, the data's integrity
	cannot always be ensured. Hence, an imperfect best-effort approach is sometimes required.

    \item \textbf{Amount of Data}: Obviously, an important goal of data acquisition is to
	extract a comprehensive data set. The best-case scenario is a bitwise copy of the entire
	memory device. Not only would we be able to analyze all the files of the user and
	the OS, but we could also restore deleted data without having to worry about data alterations.
	With mobile devices, however, this is not easily possible due to the integrated
	hardware, which necessitates intrusive methods, and the widespread use of encryption.
	It is also possible to create a bitwise copy from within the host OS which, however, 
    requires OS access and the necessary permissions.

    \item \textbf{Repeatability}: Another key factor, which needs to be considered, 
	is the repeatability of the process. Especially with mobile devices, where
	no software-based access is possible, hardware-based methods are a viable option. These,
	however, can be heavily intrusive and might involve, besides the dismantling of the
	device, the usage of soldering, or even the removal of the flash memory chip.
	By executing such processes, the device or flash memory is likely to be damaged, which
	might lead to data loss. Therefore, utilization of such methods needs to be considered
	carefully.

    \item \textbf{Ease-of-Use}: Time is always an important factor in an investigation, and even
	if a method is theoretically possible, its practical execution might entail a lot of
	effort. Gaining privileged access to the OS of a mobile device, for example, might include
	reversing and exploiting vulnerabilities, which is often very time-consuming and
	probably out-of-scope for most investigations. This obviously heavily depends on
	the situation. Sometimes screenshots of a chat history might be sufficient, and other
	times a detailed analysis of a database file or restoring deleted files are essential
	to solving the case.
\end{itemize}

For mobile forensics, most of those criteria have a conflict of interest. High data integrity is
ensured when we directly access the storage medium, which for mobile devices means possibly destructive
acquisition methods that are not easy to execute. Copying data from the OS might be simple,
but the amount of data could be limited, and the integrity cannot be ensured.
Therefore, a sufficient compromise has to be found, where we try to take all those points into account.

Concerning the amount of extractable data, three categories of data acquisition for mobile devices,
as described by~\citet{casey2011digital}, are differentiated. The simplest way of gathering
data is \emph{manual acquisition}, where the device is used as intended and important
information is preserved by screenshots. For the \emph{logical acquisition}, some form of
interaction with the file system, usually by using the OS interface, is necessary.
The simplest form can be a publicly accessible directory with shared data when the device is
plugged into a PC (e.g., the camera directory of smartphones) up to a full file system copy when
root access to the device's OS is possible. The \emph{physical acquisition} is the most complete
form of data acquisition, since a bitwise copy of the memory device is created. For
mobile devices, such an image can either be extracted from within the OS or by directly accessing
the flash memory through hardware-based methods.

By analyzing this information, we define a data acquisition sequence that considers all
criteria as well as possible. 
It is sequentially executed for both Nyon devices in the order of the following subsections until
we acquire a sufficient amount of data.

\subsection{Manual Acquisition}
\label{sec:method:manual}

First, we used manual acquisition to evaluate how much information
can be accessed using the capabilities of the user interface. This process
is very easy to use and easily repeatable but might alter data on the device, which is why
measures such as disabling the Internet connection and thoroughly documenting the process
are essential. Since we did not expect to acquire a lot of information from this process, at least
one other acquisition method is executed to obtain files from the device.

\subsection{OS-based Acquisition}
\label{sec:method:os}

Afterwards, we tried to gain access to the OS of the device, which would enable us to copy
data from the file system. If we were able to also acquire root privileges, both logical and 
physical acquisition could be executed from within the OS. We call this step
\emph{OS-based acquisition}, which has the advantage of possibly gaining access to the
entire data of the device with minimal risk of
damage or data loss. The process of gaining access to the OS, however, may vary
in complexity depending on the method used. Additionally, data
integrity cannot be guaranteed, because the OS might be compromised, or background
processes may concurrently change data, especially when an Internet or Bluetooth
connection is established.

\subsection{Hardware-based Acquisition}
\label{sec:method:hardware}

If an OS-based acquisition was not possible, direct hardware
access is considered. As a physical acquisition technique, hardware-based acquisition has the advantage of
providing us with a complete data set. 
The integrity of the acquired data is also very high since we do not rely on the OS of the
device. However, those methods require special hardware tools as well as knowledge and practice
to minimize the risk of damage.
Therefore, we first try to use debug interfaces to access the memory chip's data, 
before we actually consider the more destructive chip-off technique.


\section{Forensic Analysis of the Nyon 2014}
\label{sec:nyon14}

In this section, we forensically analyze the \SI{8}{\giga\byte} version (``BUI275'') 
of the first-generation Nyon computer (cf. \Cref{sec:nyon14basics}). 
We start by evaluating a suitable acquisition method for the Nyon~2014 
(cf. \Cref{sec:method}). 
Then, we analyze the acquired data to identify forensically relevant traces, 
and evaluate the possibility of data tampering. 

\subsection{Data Acquisition}
\label{sec:nyon14:acquisition}

The possibilities for data acquisition based on connectivity and hardware features 
of the Nyon~2014 are discussed in the following and follow the data acquisition sequence 
presented in the Sections~\ref{sec:method:manual}--\ref{sec:method:hardware}. 
Finally, the method deemed most suitable for data acquisition is executed 
and the amount of extractable data is described in detail.

\subsubsection{Manual Acquisition}
\label{sec:nyon14:acquisition:manual}

\begin{figure}[tb]
    \centering
    \includegraphics[width=0.48\textwidth]{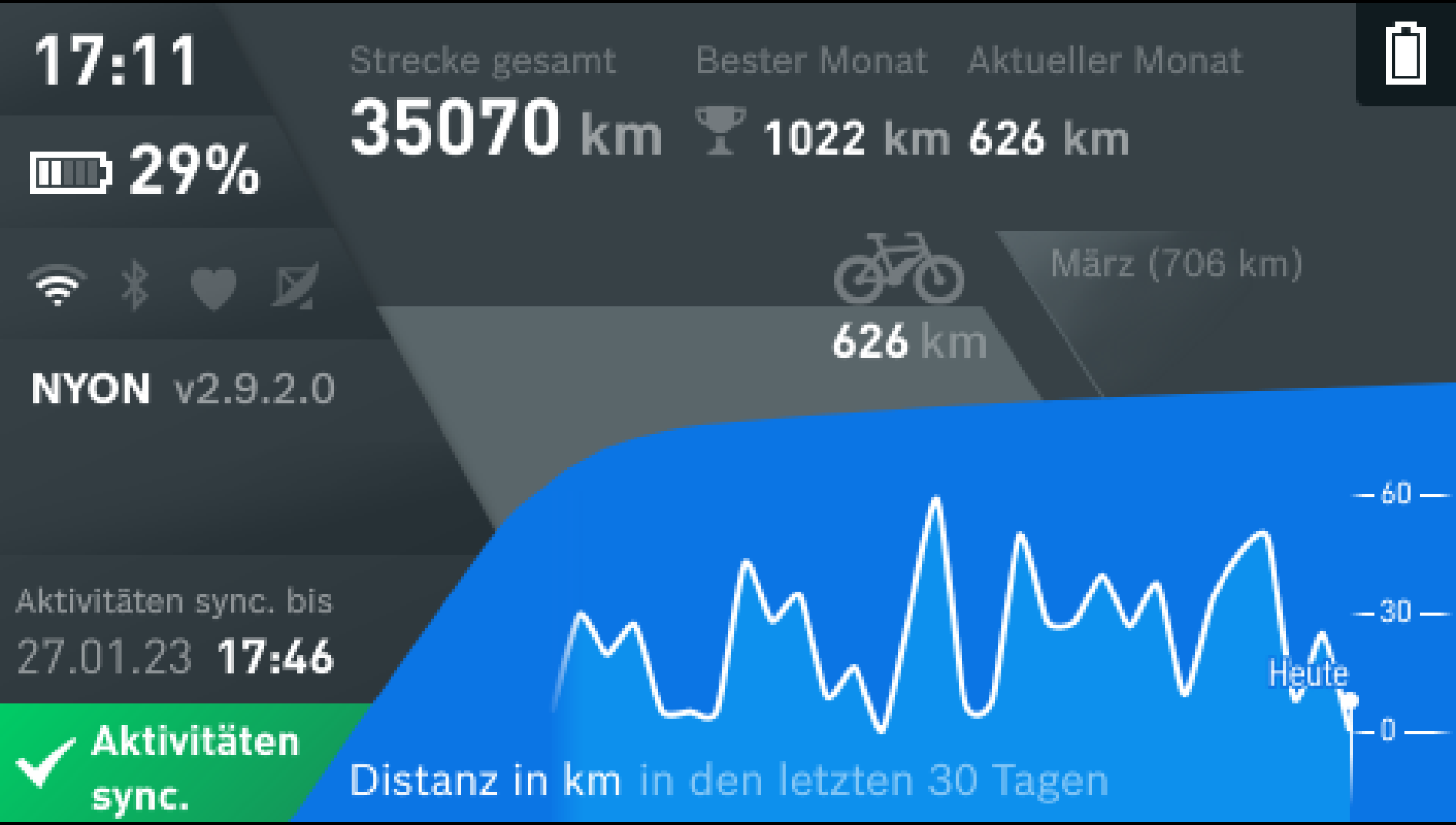}
    \caption{Dashboard of Nyon~2014.}
    \label{fig:dash1}
\end{figure}

The information provided by the user interface of the Nyon~2014 is limited. 
The dashboard depicted in \autoref{fig:dash1} shows 
a general overview of the user's monthly cycling statistics  
as well as the software version and the last synchronization with 
the Bosch cloud. 

In the settings, the user can be identified by name and email address, 
but no other personal information is disclosed. 
Further, the SSIDs of the wireless networks the Nyon~2014 was connected to are accessible. 
In the navigation screen, recent destinations are available which, however, 
are lacking respective timestamps. 
This already concludes the amount of data obtainable by manual acquisition. 

\subsubsection{OS-based Acquisition}
\label{sec:nyon14:acquisition:os}

When connecting the Nyon~2014 to a PC via USB, 
accessing the device's file system is not possible, 
but Bosch's official ``DiagnosticTool'' is able to communicate with the device. 
The software enables certified bike shops and manufacturers to debug and update 
Nyon computers. It can be freely downloaded by anyone \citep{bosch_tool} but
requires a hardware token to run. 
The most interesting feature of Bosch's DiagnosticTool is the update process 
as it might reveal internals about the Nyon device, 
and altering the process might enable access to the OS. 
Hence, we analyzed the update process in further detail.

\paragraph{Update Process.} 

\begin{figure}[tb]
    \includegraphics[width=0.5\textwidth]{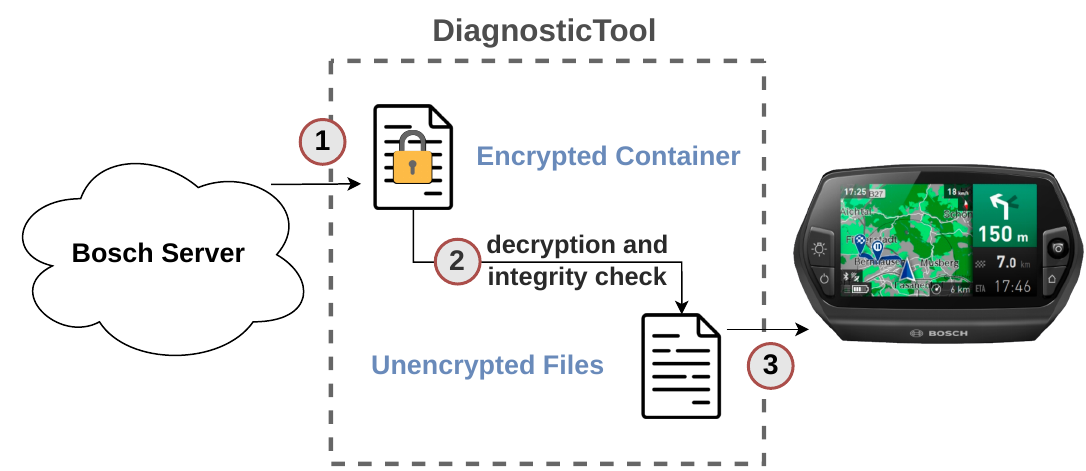}
    \caption{Update process of the Nyon~2014.}
    \label{fig:update}
\end{figure}

Through black-box analysis, we determined that the update process 
basically consists of three steps which are illustrated in \autoref{fig:update}: 

\begin{enumerate}
\item The DiagnosticTool either (i) establishes a connection to Bosch servers to download the
latest update container file, or (ii) offers to choose a local file. 
The update container file with the file extension \texttt{cff2} turned out to
be a ZIP-compressed archive. By unzipping, we uncovered a \texttt{meta.xml} file that
revealed the signing of the update container by Bosch. Furthermore, important parts of the
extracted files were encrypted.\label{item:updateone}
\item After the update container file was selected, the DiagnosticTool indicates to ``prepare flash data'', 
which seems to be the extraction and decryption process that takes place entirely on the PC running
the software. \label{item:updatetwo}
\item The last step is a reboot of the Nyon computer before the decrypted firmware
files are transferred and installed, followed by a second reboot which concludes the
process.\label{item:updatethree}
\end{enumerate}

\paragraph{Exploiting a Design Flaw in Update Process.} 

Through signing and encryption, 
the update process hinders us from forging our own update package. 
However, due to the fact that the decryption and integrity check take place on the PC and not on the Nyon, 
we were able to interfere with the process as follows to enable Telnet access: 
\begin{enumerate}

\item By letting the program run in a sandbox which stores all files
  of the process locally, we gained access to the decrypted firmware files.

\item Through analysis of the update data, the Nyon's initialization
  script (\texttt{start.sh}) that executes software on startup was identified. 
  We altered the script to start a Telnet server on the Nyon device. 

\item The timeframe between steps~\ref{item:updatetwo}
  and~\ref{item:updatethree} of the update process was utilized to
  exchange the original files with our forged ones. Therefore, we
  bypassed the integrity check and the decryption procedure, which is how
  our forged data was copied onto the device as a legitimate update.

\end{enumerate}
This design flaw in the update process gave us root access to the Nyon~2014  
using the Telnet protocol.
To easily acquire the data, we started an FTP server on the device 
which enabled us to browse and copy data from the file system.

\paragraph{Forensic Soundness.} 

Naturally, the changes applied to a system when installing an update 
might alter or delete forensically relevant user data or system files. 
While it is correct that parts of the OS are overwritten, 
the user data is usually not affected. 
This also applies in our case:

We discovered that the execution of the update process
is accomplished by the ``MFGTool'' of NXP, 
the manufacturer of the Nyon~2014's CPU (cf. \Cref{sec:nyon14basics}). 
This tool requires an XML file with instructions on how the process is executed. 
By inspecting the \path{ucl2_update_full.xml} file, 
we found that the forensically relevant user partition remains mostly untouched, 
except for two system directories that are temporarily stored on this partition 
to be restored after the system update: 
\path{/var/lib/connman/} which includes the Wi-Fi settings, and 
\path{/var/lib/bluego/} which contains information about connected Bluetooth devices. 
Hence, the user data is left unchanged, 
while the system partition is completely rewritten.

\subsubsection{Hardware-based Acquisition}
\label{sec:nyon14:acquisition:hardware}

While being forensically sound, hardware-based acquisition 
usually entails the risk of damaging the hardware  
which might lead to an unusable device or even data loss. 
Further, on devices with encrypted user data, 
hardware-based acquisition is only practical 
if the encryption key can also be obtained. 
Our analysis showed no signs of storage encryption on the Nyon~2014, 
making hardware-based acquisition generally conceivable. 
But since less intrusive OS-based acquisition was possible, 
no hardware-based methods were executed. 
For cases, where non-invasive acquisition is not applicable, 
as for the Nyon~2021, 
we present a hardware-based acquisition in \Cref{sec:nyon21:acquisition:hardware}.

\subsection{Data Evaluation}

After obtaining root access via Telnet, we were able to browse the entire file system 
of the Nyon~2014 to identify forensically relevant data. 
Unsurprisingly, despite being a rather exotic distribution, 
the file system structure of ``Poky'' (cf. \Cref{sec:nyon14basics}) 
follows the common Linux file system hierarchy.

\paragraph{Main Application.} 

We found that a single executable called \texttt{Main}---stored in a directory of the same name---%
is responsible for all the functionalities and the user interface of the Nyon~2014. 
The binaries folder and its files are located in the user's home directory
(\path{/home/appdata/}) which also contains most of the artifacts described below.

\paragraph{User.} 

Personal information about the user is contained 
in the \path{Main/Apps/Settings/<USERID>/userObject.json} file 
located in the home directory. 
An excerpt of the JSON file is shown in \autoref{lst:userobject} which reveals, 
for instance, the user's name, email address, gender, date of birth, 
address information, and social media accounts.

\begin{lstlisting}[
    language=json, 
    captionpos=b, 
    basicstyle=\small\ttfamily, 
    caption={\texttt{userObject.json} file contents}, 
    label={lst:userobject}
]
{
    "user_id":              "1234567890123",
    "date_of_birth":        "2000-01-01",
    "first_name":           "Jane",
    "last_name":            "Doe",
    "gender":               "female",
    "email":                "janedoe@example.com",
    "home_address":         { [...] },
    "mobile_phone_number":  "[...]",
    "facebook":             null,
    "twitter":              null,
    [...]
}
\end{lstlisting}

\paragraph{Cycling Statistics.} 

The SQLite database file \texttt{EBikeCharts} 
contains a single table named \textbf{ChartsData} 
whose purpose is seemingly  
to store statistical data for diagrams. 
However, we only found occurrences of the file name in log files (yet). 
Each row is associated with a ``UserId'' (cf. \autoref{lst:userobject}). 
The other columns store information related to 
the trip (``NTDistance'', ``Altitude''), 
the cyclist (``Speed'', ``driverCadence'', ``heartRate''), 
and the battery (``stateOfCharge'', ``Consumption'', ``Power''). 
As no timestamps are indicated in the table, the precise logging interval is unknown. 
However, since the values in the ``NTDistance'' column increase by steps of 25 
(e.g., 66175 $\to$ 66200), it may be cautiously assumed that 
logging occurs each 25~meters when cycling.

\paragraph{Cycling Activities.} 

Another SQLite database called \texttt{EBike} contains a total of 
seven tables whose rows are all associated with a ``UserId'' (cf. \autoref{lst:userobject}) 
and ``TimeStamp'' (epoch):

The table \textbf{Activities} stores information related to, for instance, 
trips (e.g., ``StopTime'', ``Distance''), 
hardware components of the bike (e.g., ``DriveUnitSerial'', ``BatteryPackSerials''), 
or fitness and driving performance (e.g., ``Calories'', ``Max. Speed''). %
The ambient temperature and air pressure are logged in the table \textbf{AmbientData}, 
and the table \textbf{BikeBattery} provides historical data on the battery's ``stateOfCharge''. %
Regarding the electric drive of the eBike, the table \textbf{DriveUnit} 
logs whether it ``isMoving'', the status of its ``odometer'', 
the present bike speed, and the associated torque, revolution, and power of the motor. %
Additionally, the table \textbf{Operational} logs whether the ``BUI'' 
(cf. product name in \Cref{sec:nyon14basics}) 
was operational at given timestamps. %
Driver-related information can be found in the table \textbf{Driver}, 
such as ``heartRate'', ``driverCadence'', ``driverTorque'', or ``driverPower''. %
Finally, the table \textbf{Localization} contains 
the GPS coordinates of recent cycling activities 
indicated by latitude and longitude, as well as the corresponding ``SensorAltitude''. 
We noticed that trip data is deleted from the table  
as soon as it is synchronized with the Bosch cloud. 
In practice, the table may therefore contain geodata 
of multiple past trips if no synchronization occurred, 
or none at all.

\paragraph{Planned Trips.} 

GPX files are stored in the folder \path{Main/Apps/Settings/<USERID>/gpx/}
within the home directory. The parent directory is named after the user's ID (cf. \autoref{lst:userobject}). 
These ``GPS Exchange Format'' files contain the waypoints of routes 
the user planned using the companion app or website (cf. \Cref{sec:nyon:appweb}), 
indicated by longitude and latitude values in an XML schema.

\paragraph{Bike Information.} 

The \texttt{Settings.ini} file in \path{Main/Apps/Settings/<USERID>/} 
stores technical details about the user's eBike,  
such as part and serial numbers of components, 
or the Wi-Fi access token. Furthermore, the file includes 
timestamps that indicate at which point in time the recording of fitness and geo data were allowed, 
and the date of the last sync (encoded as Qt \texttt{@Variant} string).

\paragraph{Connectivity.} 

As mentioned in \Cref{sec:nyon14:acquisition:os}, 
the two directories \path{/var/lib/connman/} and \path{/var/lib/bluego/} 
were saved and restored during the update process. 
The former contains a subdirectory for each Wi-Fi connection that has been established, 
including a \texttt{settings} file revealing the SSID, passphrase, network settings, and timestamp.
The latter stores a file for every Bluetooth device, 
including general information and the device's name.

\paragraph{System Logs.} 

Detailed system information can be found in the system's log files, 
which are located in the \path{/home/appdata/var/log/} directory. 
The scope of these log files covers, for instance, 
Bluetooth and Wi-Fi, system messages, USB diagnostics, and shutdown events.

\paragraph{Tampering.} 

In principle, if obtaining unrestricted access to the OS like we did (cf. \Cref{sec:nyon14:acquisition:os}), 
data can be manipulated by anybody with physical access to the Nyon~2014. 
In an experiment, we forged a trip by adding additional 
GPS points and timestamps to the \texttt{EBike} database file. 
As expected, the trip was accepted by the device without any problems. 
Moreover, it was also synchronized with Bosch's cloud 
and thereby displayed in both Bosch's companion app and web service (cf. \Cref{sec:nyon:appweb}). 
Even though it might take some effort to forge a realistic trip 
with, for instance, reasonable altitude and distance values as well as timestamps, 
tampering with data on the Nyon~2014 is feasible if enough time and expertise is given.


\section{Forensic Analysis of the Nyon 2021}
\label{sec:nyon21}

We now analyze the Nyon~2021, the second-generation device in the series of
bike computers that is still the most recent Nyon at the time of writing. 
Again, the acquisition methods presented in the Sections~\ref{sec:method:manual}--\ref{sec:method:hardware} 
are evaluated, whereby the most suitable method is executed. 
Afterwards, the acquired data is analyzed to identify relevant traces.

\subsection{Data Acquisition}
\label{sec:nyon21:acquisition}

According to our acquisition methodology, we first evaluate manual acquisition, 
where we highlight the differences and similarities to the Nyon~2014 (cf. \Cref{sec:nyon14:acquisition:manual})  
before trying to acquire the actual data from the device.
For the first-generation Nyon, we were able to extract the data logically 
by using a design flaw in the update process (cf. \Cref{sec:nyon14:acquisition:os}), 
which had a low footprint and enabled us to acquire the entire data set without
risking damage to the device.  
Therefore, OS-based acquisition is considered for the Nyon~2021 as well, 
before the more intrusive hardware-based acquisition is evaluated.

\subsubsection{Manual Acquisition}
\label{sec:nyon21:acquisition:manual}

Similar to the first-generation Nyon (cf. \Cref{sec:nyon14:acquisition:manual}), 
not much information can be gained through the user interface of the Nyon~2021: 
The settings only provide the user's name and email address, 
and the navigation screen only displays recently typed addresses. 
Regarding the latter, neither a timestamp nor an indication on whether those destinations 
were actually visited or only searched is visible. 

A minor difference to the Nyon~2014 is that only destinations are displayed  
which have been typed in by the user or have been synchronized from the companion app or web service. 
Hence, the Nyon~2021 does not display destinations that were started from the app directly.

\subsubsection{OS-based Acquisition}
\label{sec:nyon21:acquisition:os}

Similar to its predecessor, the Nyon~2021 can be connected to a PC 
for debugging purposes or installing updates via Bosch's DiagnosticTool. 
In contrast to the Nyon~2014, however, 
we were not able to interfere with the update process anymore. 
Apparently, the design flaw reported in \Cref{sec:nyon14:acquisition:os} 
was recognized and fixed by Bosch for the second-generation Nyon. 
In the revised update process, the Nyon~2021 is now being recognized as a storage medium on the PC, 
and the DiagnosticTool just copies over an encrypted update file. 
This file contains the encrypted file system image (``SquashFS'')  
which gets decrypted on the device itself, leaving us without a race-window in the process.

As known from Android devices, we identified the debug mode of the Nyon~2021  
by clicking 17~times on ``SW-Version'' in the settings menu. 
After activating the switch labeled ``Debug'', the Nyon~2021 was connected to a PC 
and recognized as a ``RNDIS'' device, i.e., as a USB network device  
(``Remote Network Driver Interface Specification''). 
Accordingly, a new network device with the IP address ``172.16.35.101''  
and an open port ``5001'' was recognized. 
In Windows, the new device is called ``Bosch Service Bridge'', which seems to be a debug service.
The AppArmor file for the ``Diagnosticsservices'' binary, 
which seems to be the counterpart on the device, 
suggests that it can be used to spawn a shell. 
Since the usage of this mechanism for data acquisition 
would require time-consuming reverse engineering
and possibly exploitation, the feasibility remains unknown. 

In summary, the OS-based acquisition revealed no possibility to access the Nyon~2021. 
In accordance with our acquisition methodology, 
we therefore opted for a more intrusive hardware-based acquisition, as explained below.

\subsubsection{Hardware-based Acquisition}
\label{sec:nyon21:acquisition:hardware}

As a last resort, hardware-based methods were executed
to acquire data from the Nyon~2021. Since it is the least destructive solution, 
we first attempted access over potential hardware debug interfaces, 
followed by a destructive chip-off procedure to obtain and read the eMMC chip.

Active JTAG or UART debug interfaces are useful 
to obtain data from IoT devices \citep{gordon2019hardware}. 
In our case, a possible hardware debug interface was identified on the MLB 
and highlighted in \Cref{fig:nyon2board}. 
We soldered wires to the pre-tinned pads, 
and connected them to a logic analyzer and JTAGulator. 
As the pads only showed either constant or no supply voltages, 
this approach led to no result. 

After all non-destructive approaches in our methodology were exhausted, 
we executed a chip-off. The basic idea of this method is to remove a chip 
from the printed circuit board (PCB) by liquifying the solder through heat, 
and extracting the data using special adapters suitable for the respective chip type. 
However, as described by~\citet{fukami2017improving},
this procedure might lead to data loss because of the high temperatures that the chip is exposed
to, and surrounding components might also be damaged. 
Therefore, a reassembly of the device to a working condition is unlikely.

\begin{figure}[tb]
    \centering
    \begin{subfigure}[b]{0.155\textwidth}
	    \includegraphics[width=\textwidth]{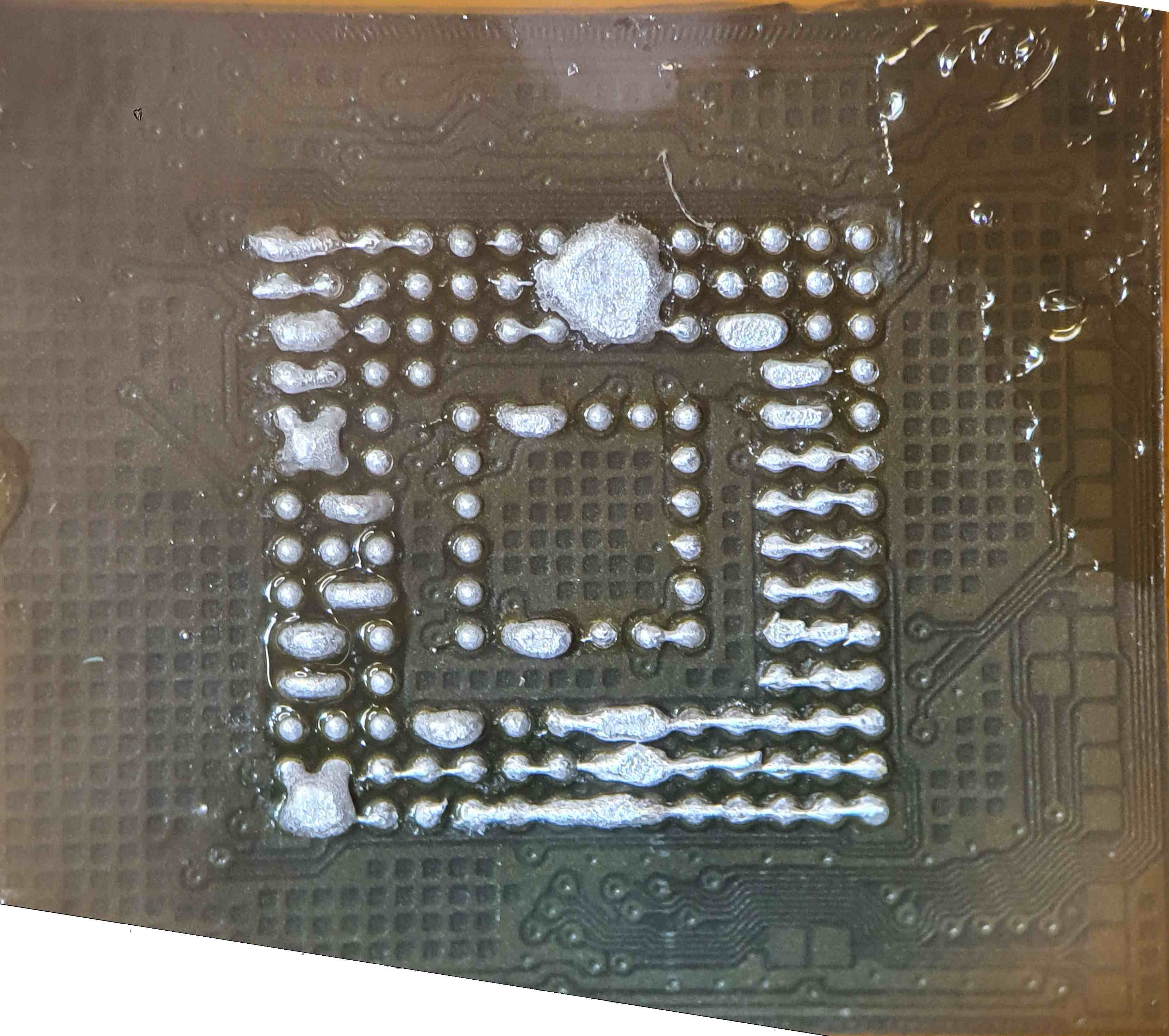}
	    \caption{After removal}
	    \label{fig:emmc1}
    \end{subfigure}
    \hfill
    \begin{subfigure}[b]{0.155\textwidth}
	    \includegraphics[width=\textwidth]{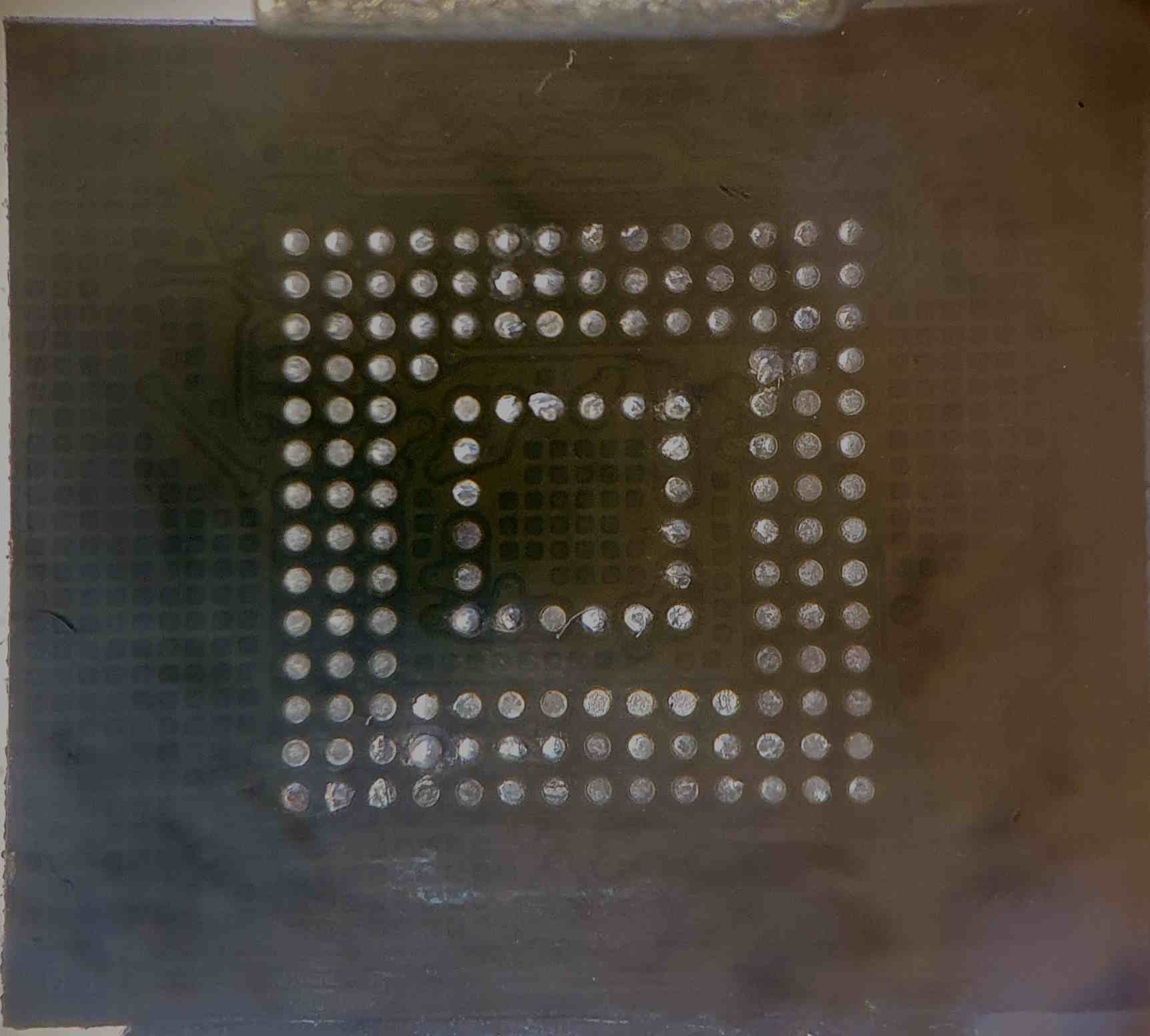}
	    \caption{Cleaned}
	    \label{fig:emmc2}
    \end{subfigure}
    \hfill
    \begin{subfigure}[b]{0.155\textwidth}
	    \includegraphics[width=\textwidth]{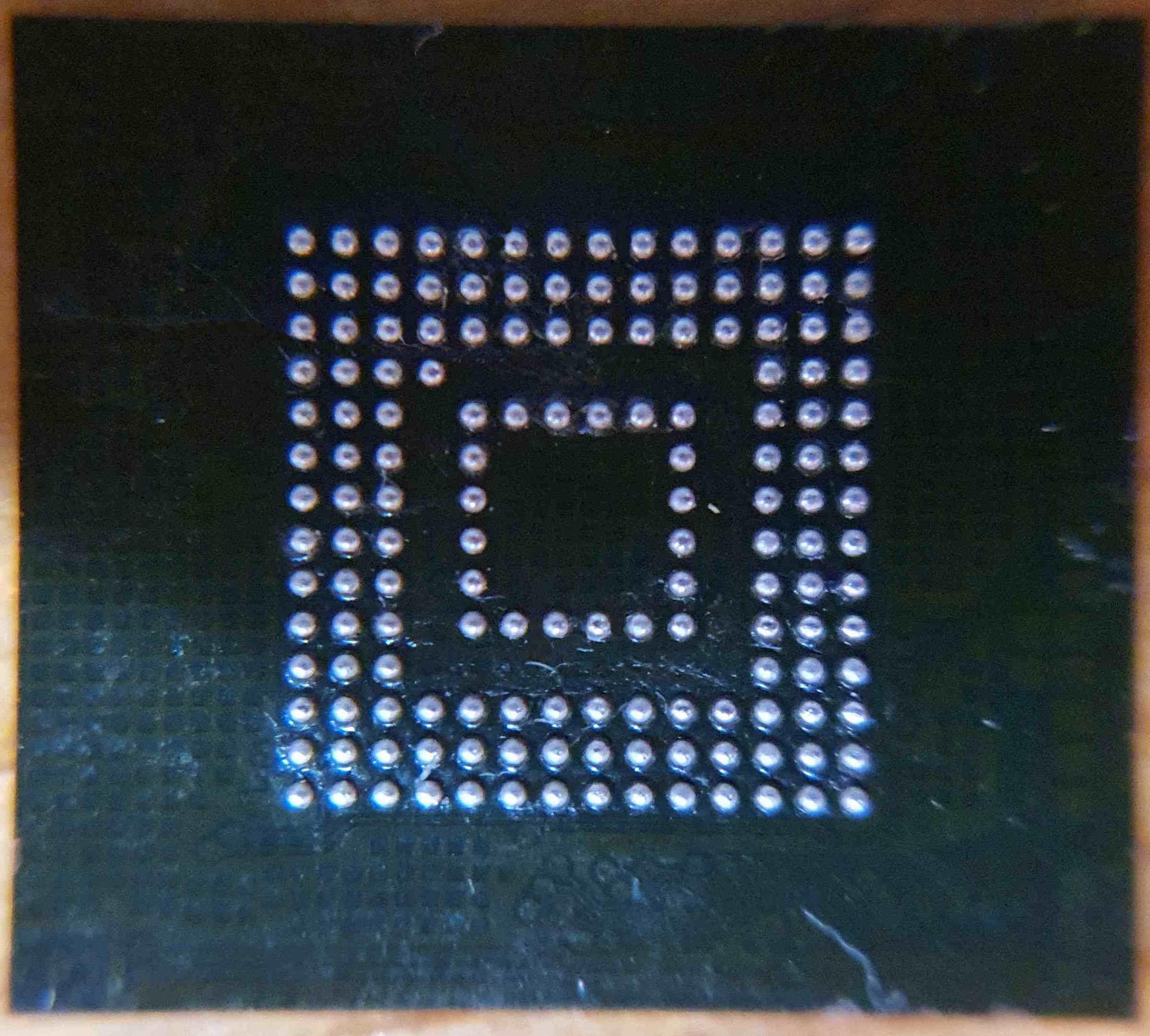}
	    \caption{Reballed}
	    \label{fig:emmc3}
    \end{subfigure}
    \caption{The removed eMMC chip after the chip-off process.}
    \label{fig:emmc}
\end{figure}

In \autoref{fig:emmc}, we show the various stages of the eMMC throughout the chip-off procedure,  
which were as follows:

\begin{enumerate}
\item Before heating the chip using a hot air station, 
we placed heat-resistant tape around the edges and over the
surrounding components to reduce damage. 
Afterwards, the chip was preheated to \SI{250}{\celsius} for about 3~minutes. 
Generally speaking, the temperature set on the hot air station, 
the temperature arriving at the chip package, 
and the temperature reaching the solder underneath the chip, are not identical. 
Then, while continuously adding flux, the temperature
was increased to \SI{380}{\celsius} until the solder liquified, 
and the chip could be removed. 
As shown in \autoref{fig:emmc1}, the chip was covered with
excessive solder and flux. 
\item The chip was cleaned using a soldering iron and desoldering braid. 
Contaminants were removed using isopropyl alcohol. 
The result can be seen in~\autoref{fig:emmc2}. 
\item We reballed the chip using a stencil for the
respective BGA-153 socket, soldering paste, and hot air. 
The result is shown in~\autoref{fig:emmc3}.
\item Using the Easy JTAG Plus (2.53 rev.\,2), 
an eMMC adapter for the BGA-153 socket, 
and the software ``EasyJtag Classic Suite'' (v3.7.0.24), 
we successfully obtained access to the data on the chip. 
In total, 8 partitions were extracted, as well as unallocated memory. 
\end{enumerate}

\subsection{Data Evaluation}

The most important of the extracted partitions are described in~\autoref{tab:partitions}. 
Besides the boot image, the first partition contained files with partition
names, which stored their offset, size, hash, and salt values. Those files are
probably used for the ``dm-verity'' feature which checks the consistency of important
system partitions on startup. The names that are provided in the table, 
helped to attribute the partitions to their purposes. 
Most of the partitions are not forensically relevant, 
since they cannot change due to ``dm-verity''.

\begin{table*}[tb]
    \rmfamily
    \begin{tabular}{@{}p{2.4cm} r l p{10.0cm}@{}}
    \toprule
    \textbf{START} & \textbf{SIZE} & \textbf{NAME} & \textbf{INFO} \\
    \midrule
    0x400000 & \SI{16}{\mega\byte} & - & contains files with partitioning scheme and partition names \\
    0x2800000 & \SI{1}{\giga\byte} & \texttt{bui3xx-image} & system partition with the root file system \\
    0x42900000 & \SI{192}{\mega\byte} & \texttt{bui3xx-systemconfig} & log files and system config \\
    0x4ea00000 & \SI{336}{\mega\byte} & \texttt{bui3xx-recovery} & system recovery partition \\
    0x6ac00000 & \SI{5}{\giga\byte} & - & skobbler map data for navigation, key for userdata partition \\
    0x1ac900000 & \SI{50}{\mega\byte} & - & contains test images and key material \\
    0x1af900200 & \SI{550}{\mega\byte} & - & unallocated memory, turned out to be encrypted userdata partition \\\bottomrule
    \end{tabular}
    \caption{Important partitions with their offset, size, names (according to the partition description files), and content information.}
    \label{tab:partitions}
\end{table*}

\paragraph{Userdata Partition.}

The unallocated memory extracted from the eMMC turned out 
to be the LUKS-encrypted userdata partition. 
The header of the respective dump file stored information 
about the key material used for encryption, which is 32-bit in size. 
Since we already identified keys on other partitions, we tried our luck and searched for 32-bit
files on all partitions. Hereby, we found a file named \texttt{crypto\_keyfile.bin} 
that was successfully used to decrypt and mount the userdata partition. 
As expected, this partition contained all the forensically relevant data 
in two directories, namely \path{system/} and \path{users/}. 
A selection is presented below.

\paragraph{Known Wireless Networks.} 

Previously connected Wi-Fi networks are stored 
in the \path{WifiManagerSettings.json} file, 
incl. their names and passwords in plain text 
(see \autoref{lst:nyon21:WifiManagerSettings}).

\begin{lstlisting}[
    basicstyle=\small\ttfamily, language=json, captionpos=b, label={lst:nyon21:WifiManagerSettings}, 
    caption={\texttt{WifiManagerSettings.json} (Excerpt, Nyon~2021).}]
{
    "id":       "\"Galaxy Note10+0c95\"",
    "psk":      "[PLAINTEXT_PASSWORD]",
    "security": "WPA2"
}
\end{lstlisting}

\paragraph{Last Known Location.} 

The last known location, indicated by latitude, longitude, altitude, speed, and timestamp, 
can be obtained from the \path{gnssSettings.json} file (see \autoref{lst:nyon21:gnssSettings}). 

\begin{lstlisting}[
    basicstyle=\small\ttfamily, language=json, captionpos=b, label={lst:nyon21:gnssSettings}, 
    caption={\texttt{gnssSettings.json} (Excerpt, Nyon~2021).}]
"lastPosition":{
    "latitude":     [REDACTED],
    "longitude":    [REDACTED],
    "altitude":     311.0,
    "speed":        0.32667222619056702,
    "timestamp":    1686737311649,
    [...]}
\end{lstlisting}

\paragraph{Nearby and Trusted Bluetooth Devices.} 

A log file of the ``Chromium Embedded Framework'' (CEF), 
responsible for the user interface, 
is located in \path{system/webfs/logs/cef_debug.log} 
and reveals, among other things, information about 
discovered and trusted Bluetooth devices.

\paragraph{System Logs.}

The database \path{system/db/analytics.db} logs timestamps of various system events 
in the table \textbf{analytics\_events}, such as 
``BUI\-350\_SYSTEM\_WAKEUP'', ``BUI\-350\_BOOT\_INFO'', or ``BUI\-350\_START\_NAVIGATION''. 
If available, related parameters are indicated as well 
(e.g., ``\{"duration": "12:36"\}'' for ``BUI\-350\_TRIP\_\-RESET'').

The directory \path{system/webfs/logs/log/} contains several log files 
related to system activities, such as \path{system.log} or \path{wifi.log}. 
Additionally, the names of certain log files indicate software that is actively 
running on the Nyon~2021, such as ``Tunnelblick'' (OpenVPN), 
``CUPS'' (Common Unix Printing System), 
``Adobe Acrobat Updater'', ``fsck'', or ``nginx''.

\paragraph{User.}

All relevant user-specific data can be found in the \path{users/buiowner/data/system/db/} directory. 
Similar to the JSON file of the first Nyon generation (\texttt{userObject.json}), 
the \texttt{active-account.json} file contains user information, like user ID, name, address,
and contact info.

In \texttt{user-settings.db}, a total of seven tables exist, but only \textbf{settings\_app} and
\textbf{settings\_system} contain data. Judging by the values in the ``stored\_setting'' column, 
the former table contains different information for the application, 
such as the battery level, or whether the developer mode was activated (cf. \Cref{sec:nyon21:acquisition:os}). 
The latter table, on the other hand, provides system settings 
like the language and unit of measurement, but also user-defined goals, 
such as weekly kilometers, or monthly calories.

\paragraph{Bike Information.} 

The \texttt{bike-info.json} file contains a list of all registered eBikes, including
their serial and part numbers, software and hardware version numbers, 
as well as detailed information about their battery packs.

\paragraph{Cycling Activities.}  

The database file \texttt{NavStorage.sqlite} stores navigation and
location data of the user. 
While the \textbf{Consumptions} table stores vectors indicating the energy consumption 
for the travel distance calculation, the \textbf{Locations} table stores user-defined places 
indicated by their address (``name''), a tuple of latitude and longitude values, 
and an epoch timestamp of the last modification. 
Indicated similarly, the planned routes of users can be found in the \textbf{Routes} table, 
and recent destinations in the \textbf{Recents} table. 

The contents of another database file, \texttt{tracking.db}, 
are similar to the \texttt{EBike} database of the first-generation Nyon, 
even though they are now spread over 22~tables. 
More interestingly, we found that data synchronization on the Nyon~2021 
seems to not have an impact on the stored data. Contrary to the Nyon~2014, 
where the data was deleted after synchronization with the Bosch cloud, 
the second-generation Nyon stored the last 100~trips.


\section{Discussion}
\label{sec:implications}

In general, the data we could acquire on both generations of Nyons was similar, albeit the
data set on the latest device was slightly larger and the structure differed.

\paragraph{Forensic Implications.}
Probably the most obvious forensically interesting data is the information about the trips taken
by the user. This information is available comprehensively on both devices, even though trips on
the first Nyon are deleted after a synchronization.
In particular, the wealth of detailed location data and timestamps is of
utmost importance for forensics. In a hypothetical case, it might help to determine if a suspect
visited a certain location, and at which time the visit took place.

Furthermore, both devices feature detailed information about the driving behavior of the user.
While not being as relevant as the location information, it could help to determine the plausibility
of the assumption that a certain person was actually the one who drove this bike.
If, for example, a physically demanding cycling style is exhibited, and a fairly non-athletic person
is the suspected driver, it could be an indication of a false assumption.

Personal information about the user could also be of interest since not only the
name but also contact info and social media accounts are stored. If this data
was not previously known or not disclosed, it might be of assistance
in drawing a comprehensive picture of a suspect.

The data about wireless networks and Bluetooth devices within proximity can
enhance our confidence in ascertaining the locations visited by the device's user. Of particular
significance are nearby Bluetooth devices, which can potentially aid in determining
whether an individual was actually the driver based on the registration of their phone as
a Bluetooth device. Additionally, the examination of stored Wi-Fi passwords offer the option
for network analysis.

Due to the unrestricted access to the OS of the first Nyon, we showed that data tampering is
possible, which should be considered when interpreting traces. 
For example, a trip does not have to be completely made up to create a false alibi, because changing
the timestamps might be enough.

\paragraph{Advice for Practitioners.}
From working on this project, we especially noticed many similarities between
smartphones and the Nyons.
However, some differences require distinct data acquisition methods.
Since the security measures were not as advanced, hardware-based data acquisition 
considered outdated for modern smartphones retained their applicability. 
The Nyons also do not offer publicly accessible OS interfaces usable for data acquisition, such as shared
folders, local backups, or publicly available debug tools.

Chip-off, as a well-known hardware-based option, is applicable on both devices,
since the Nyon~2014's data was unencrypted, and the encryption key of the Nyon 2021 was easily accessible.
But before actually executing such a destructive method, other options
should be considered.

On the software side, manual gathering of data from the user interface did not result in a comprehensive data set,
however in some cases this information might suffice.
As we could show for the Nyon~2014, a thorough analysis of possible software-based access methods might be worth the
effort. Particularly so, since it became possible to avoid more destructive options and
evidence about the data's integrity due to the processes execution could be provided.


\section{Conclusion and Future Work}
\label{sec:conclusion}

We conducted a comprehensive forensic analysis of both existing Bosch Nyon eBike board computers
to uncover valuable digital traces with forensic significance. Before acquiring the data,
we examined both the hardware and software aspects of the devices. This involved highlighting the
essential features by inspecting the main board with its hardware components and interfaces, 
as well as by analyzing software details and communication protocols.

Subsequently, we formulated a structured data acquisition methodology by first defining important
forensic priorities concerning data acquisition and their challenges in the field of
mobile forensics. By taking those requirements into account, a structured acquisition strategy
was formulated.

Our efforts led to the successful data acquisition for both generations of
Nyon eBike computers. For the first-generation Nyon, we acquired root access to the device's operating
system, granting us comprehensive access to its data. In contrast, the increased security measures of the
second-generation Nyon necessitated using hardware-based methods for data extraction.
Due to our inability to access the device through hardware debug interfaces, we resorted to a chip-off
procedure, which enabled the extraction of the data from the flash memory. The extracted user data 
was encrypted, but since the encryption key was stored on an unencrypted partition, decryption was possible.

By analyzing the files from both Nyon devices, we uncovered a wealth of forensically relevant
information. While many similarities existed between the data of the two generations, the second-generation Nyon
had a slightly broader spectrum of data. On both devices, detailed location records were available, including
timestamps, driving behavior profiles, stored Wi-Fi credentials, Bluetooth logs,
and system data. The most apparent applicability of this information are statements about a suspect's
whereabouts and assessments about the plausibility of the individual's driving behavior.

In conclusion, our forensic analysis of Bosch Nyon eBike computers not only improves our
understanding of these special-purpose mobile devices but also highlights their significance in
the forensic context. By addressing the inherent challenges and successfully extracting
valuable data, our study contributes to the evolving landscape of digital forensics,
underscoring the importance of considering such devices in the investigative process.

A possible future direction is the analysis of the app and cloud data from the Nyon devices,
as well as an evaluation of a broader spectrum of eBike board computers.

\medskip

The results of this paper have been reported to Bosch and have undergone a coordinated disclosure process with Bosch PSIRT.

\subsection*{Acknowledgements}
We thank Jenny Ottmann as well as the anonymous reviewers for their helpful comments. %
This work has been supported partially by the Bavarian Ministry of Science and Arts 
as part of the project ``Security in Everyday Digitization'' (ForDaySec) %
and by Deutsche
Forschungsgemeinschaft (DFG, German Research Foundation) as part of
the Research and Training Group 2475 ``Cybercrime and Forensic
Computing'' (grant number 393541319/GRK2475/1-2019).

\bibliography{nyon}

\begin{thebibliography}{28}
\expandafter\ifx\csname natexlab\endcsname\relax\def\natexlab#1{#1}\fi
\providecommand{\url}[1]{\texttt{#1}}
\providecommand{\href}[2]{#2}
\providecommand{\path}[1]{#1}
\providecommand{\DOIprefix}{doi:}
\providecommand{\ArXivprefix}{arXiv:}
\providecommand{\URLprefix}{URL: }
\providecommand{\Pubmedprefix}{pmid:}
\providecommand{\doi}[1]{\href{http://dx.doi.org/#1}{\path{#1}}}
\providecommand{\Pubmed}[1]{\href{pmid:#1}{\path{#1}}}
\providecommand{\bibinfo}[2]{#2}
\ifx\xfnm\relax \def\xfnm[#1]{\unskip,\space#1}\fi
\bibitem[{Barr-Smith et~al.(2021)Barr-Smith, Farrant, Leonard-Lagarde, Rigby,
  Rigby and Sibley-Calder}]{barr2021dead}
\bibinfo{author}{Barr-Smith, F.}, \bibinfo{author}{Farrant, T.},
  \bibinfo{author}{Leonard-Lagarde, B.}, \bibinfo{author}{Rigby, D.},
  \bibinfo{author}{Rigby, S.}, \bibinfo{author}{Sibley-Calder, F.},
  \bibinfo{year}{2021}.
\newblock \bibinfo{title}{Dead man's switch: forensic autopsy of the nintendo
  switch}, in: \bibinfo{booktitle}{Proceedings of the Digital Forensics
  Research Conference Europe (DFRWS EU)}.
\bibitem[{Buquerin et~al.(2021)Buquerin, Corbett and
  Hof}]{buquerin2021generalized}
\bibinfo{author}{Buquerin, K.K.G.}, \bibinfo{author}{Corbett, C.},
  \bibinfo{author}{Hof, H.J.}, \bibinfo{year}{2021}.
\newblock \bibinfo{title}{A generalized approach to automotive forensics}, in:
  \bibinfo{booktitle}{Proceedings of the Digital Forensics Research Conference
  Europe (DFRWS EU)}.
\bibitem[{Casey(2011)}]{DBLP:books/daglib/0027500}
\bibinfo{author}{Casey, E.}, \bibinfo{year}{2011}.
\newblock \bibinfo{title}{Digital Evidence and Computer Crime - Forensic
  Science, Computers and the Internet, 3rd Edition}.
\newblock \bibinfo{publisher}{Academic Press}.
\newblock \URLprefix
  \url{http://www.elsevierdirect.com/product.jsp?isbn=9780123742681}.
\bibitem[{Casey and Turnbull(2011)}]{casey2011digital}
\bibinfo{author}{Casey, E.}, \bibinfo{author}{Turnbull, B.},
  \bibinfo{year}{2011}.
\newblock \bibinfo{title}{{Digital evidence on mobile devices}}.
\newblock \bibinfo{journal}{Digital Evidence and Computer Crime}
  \bibinfo{volume}{3}, \bibinfo{pages}{1--44}.
\bibitem[{Ebbers et~al.(2021)Ebbers, Ising, Saatjohann and
  Schinzel}]{ebbers2021grand}
\bibinfo{author}{Ebbers, S.}, \bibinfo{author}{Ising, F.},
  \bibinfo{author}{Saatjohann, C.}, \bibinfo{author}{Schinzel, S.},
  \bibinfo{year}{2021}.
\newblock \bibinfo{title}{Grand theft app: Digital forensics of vehicle
  assistant apps}, in: \bibinfo{booktitle}{Proceedings of the 16th
  International Conference on Availability, Reliability and Security}, pp.
  \bibinfo{pages}{1--6}.
\bibitem[{Freiling et~al.(2018)Freiling, Gro{\ss}, Latzo, M{\"{u}}ller and
  Palutke}]{DBLP:journals/dt/FreilingGLMP18}
\bibinfo{author}{Freiling, F.C.}, \bibinfo{author}{Gro{\ss}, T.},
  \bibinfo{author}{Latzo, T.}, \bibinfo{author}{M{\"{u}}ller, T.},
  \bibinfo{author}{Palutke, R.}, \bibinfo{year}{2018}.
\newblock \bibinfo{title}{Advances in forensic data acquisition}.
\newblock \bibinfo{journal}{{IEEE} Des. Test} \bibinfo{volume}{35},
  \bibinfo{pages}{63--74}.
\newblock \URLprefix \url{https://doi.org/10.1109/MDAT.2018.2862366},
  \DOIprefix\doi{10.1109/MDAT.2018.2862366}.
\bibitem[{Fukami et~al.(2017)Fukami, Ghose, Luo, Cai and
  Mutlu}]{fukami2017improving}
\bibinfo{author}{Fukami, A.}, \bibinfo{author}{Ghose, S.},
  \bibinfo{author}{Luo, Y.}, \bibinfo{author}{Cai, Y.}, \bibinfo{author}{Mutlu,
  O.}, \bibinfo{year}{2017}.
\newblock \bibinfo{title}{{Improving the reliability of chip-off forensic
  analysis of NAND flash memory devices}}, in: \bibinfo{booktitle}{Proceedings
  of the Digital Forensics Research Conference Europe (DFRWS EU)}.
\bibitem[{Fukami et~al.(2021)Fukami, Stoykova and Geradts}]{fukami2021new}
\bibinfo{author}{Fukami, A.}, \bibinfo{author}{Stoykova, R.},
  \bibinfo{author}{Geradts, Z.}, \bibinfo{year}{2021}.
\newblock \bibinfo{title}{{A new model for forensic data extraction from
  encrypted mobile devices}}.
\newblock \bibinfo{journal}{Forensic Science International: Digital
  Investigation} \bibinfo{volume}{38}, \bibinfo{pages}{301169}.
\bibitem[{Garfinkel(2010)}]{DBLP:journals/di/Garfinkel10}
\bibinfo{author}{Garfinkel, S.L.}, \bibinfo{year}{2010}.
\newblock \bibinfo{title}{Digital forensics research: The next 10 years}.
\newblock \bibinfo{journal}{Digit. Investig.} \bibinfo{volume}{7},
  \bibinfo{pages}{S64--S73}.
\newblock \URLprefix \url{https://doi.org/10.1016/j.diin.2010.05.009},
  \DOIprefix\doi{10.1016/j.diin.2010.05.009}.
\bibitem[{Garfinkel(2013)}]{DBLP:journals/compsec/Garfinkel13}
\bibinfo{author}{Garfinkel, S.L.}, \bibinfo{year}{2013}.
\newblock \bibinfo{title}{Digital media triage with bulk data analysis and
  bulk{\_}extractor}.
\newblock \bibinfo{journal}{Comput. Secur.} \bibinfo{volume}{32},
  \bibinfo{pages}{56--72}.
\newblock \URLprefix \url{https://doi.org/10.1016/j.cose.2012.09.011},
  \DOIprefix\doi{10.1016/j.cose.2012.09.011}.
\bibitem[{Geus et~al.(2023)Geus, Ottmann and Freiling}]{geus23backup}
\bibinfo{author}{Geus, J.}, \bibinfo{author}{Ottmann, J.},
  \bibinfo{author}{Freiling, F.}, \bibinfo{year}{2023}.
\newblock \bibinfo{title}{{Systematic Evaluation of Forensic Data Acquisition
  using Smartphone Local Backup}}, in: \bibinfo{booktitle}{Proceedings of the
  Digital Forensics Research Conference (DFRWS US)}.
\bibitem[{G{\'{o}}mez et~al.(2021)G{\'{o}}mez, Mond{\'{e}}jar, G{\'{o}}mez and
  Mart{\'{\i}}nez}]{DBLP:journals/di/GomezMGM21}
\bibinfo{author}{G{\'{o}}mez, J.M.C.}, \bibinfo{author}{Mond{\'{e}}jar, J.C.},
  \bibinfo{author}{G{\'{o}}mez, J.R.}, \bibinfo{author}{Mart{\'{\i}}nez,
  J.L.M.}, \bibinfo{year}{2021}.
\newblock \bibinfo{title}{Developing an iot forensic methodology. {A} concept
  proposal}, in: \bibinfo{booktitle}{Proceedings of the Digital Forensics
  Research Conference Europe (DFRWS EU)}.
\bibitem[{Gordon et~al.(2019)Gordon, Kilgore, Wylds and
  Nowatkowski}]{gordon2019hardware}
\bibinfo{author}{Gordon, T.}, \bibinfo{author}{Kilgore, E.},
  \bibinfo{author}{Wylds, N.}, \bibinfo{author}{Nowatkowski, M.},
  \bibinfo{year}{2019}.
\newblock \bibinfo{title}{{Hardware reverse engineering tools and techniques}},
  in: \bibinfo{booktitle}{2019 SoutheastCon}, \bibinfo{organization}{IEEE}. pp.
  \bibinfo{pages}{1--6}.
\bibitem[{Hoppe et~al.(2012)Hoppe, Kuhlmann, Kiltz and
  Dittmann}]{DBLP:conf/safecomp/HoppeKKD12}
\bibinfo{author}{Hoppe, T.}, \bibinfo{author}{Kuhlmann, S.},
  \bibinfo{author}{Kiltz, S.}, \bibinfo{author}{Dittmann, J.},
  \bibinfo{year}{2012}.
\newblock \bibinfo{title}{It-forensic automotive investigations on the example
  of route reconstruction on automotive system and communication data}, in:
  \bibinfo{editor}{Ortmeier, F.}, \bibinfo{editor}{Daniel, P.} (Eds.),
  \bibinfo{booktitle}{Computer Safety, Reliability, and Security - 31st
  International Conference, {SAFECOMP} 2012, Magdeburg, Germany, September
  25-28, 2012. Proceedings}, \bibinfo{publisher}{Springer}. pp.
  \bibinfo{pages}{125--136}.
\newblock \URLprefix \url{https://doi.org/10.1007/978-3-642-33678-2\_11},
  \DOIprefix\doi{10.1007/978-3-642-33678-2\_11}.
\bibitem[{Huang(2003)}]{xbox:2003}
\bibinfo{author}{Huang, A.}, \bibinfo{year}{2003}.
\newblock \bibinfo{title}{Hacking the Xbox: An Introduction to Reverse
  Engineering}.
\newblock \bibinfo{publisher}{No Starch Press}.
\bibitem[{{Kioxia Corp.}(2019)}]{kioxia:THGBMJG6C1LBAIL}
\bibinfo{author}{{Kioxia Corp.}}, \bibinfo{year}{2019}.
\newblock \bibinfo{title}{{Datasheet: eMMC THGBMJG6C1LBAIL (8GB)}}.
\newblock \URLprefix
  \url{https://datasheet.lcsc.com/lcsc/2004271813_KIOXIA-THGBMJG6C1LBAIL_C524518.pdf}.
  \bibinfo{note}{{Accessed: 2023-10-09}}.
\bibitem[{{Micron Technology}(2018)}]{micron:JWB18}
\bibinfo{author}{{Micron Technology}}, \bibinfo{year}{2018}.
\newblock \bibinfo{title}{{Datasheet: eMMC MTFC8GACAANA-4M IT (JWB18)}}.
\newblock \URLprefix
  \url{https://www.micron.com/products/managed-nand/emmc/part-catalog/mtfc8gacaana-4m-it}.
  \bibinfo{note}{{Accessed: 2023-10-09}}.
\bibitem[{{Nanya Technology}(2015)}]{nanya:NT5CB128M16FP_DII}
\bibinfo{author}{{Nanya Technology}}, \bibinfo{year}{2015}.
\newblock \bibinfo{title}{{Datasheet: DDR3(L) SDRAM (NT5CB128M16FP-DII)}}.
\newblock \URLprefix
  \url{https://pdf1.alldatasheet.com/datasheet-pdf/view/1132525/NANYA/NT5CB128M16FP-DII/+01__7W8XLzx/1DdSZCIIvwpZDITePuab+/datasheet.pdf}.
  \bibinfo{note}{{Accessed: 2023-10-09}}.
\bibitem[{{Nanya Technology}(2022)}]{nanya:NT5CC256N16ER_EKI}
\bibinfo{author}{{Nanya Technology}}, \bibinfo{year}{2022}.
\newblock \bibinfo{title}{{Datasheet: DDR3(L) SDRAM (NT5CC256N16ER-EKI)}}.
\newblock \URLprefix
  \url{https://static6.arrow.com/aropdfconversion/54ca98a15fba86b32611a8fd857f21d97cc3239f/4gb_ddr3_e_die_component_datasheet.pdf}.
  \bibinfo{note}{{Accessed: 2023-10-09}}.
\bibitem[{{NXP Semiconductors}(2018)}]{nxp:MCIMX6S5EVM10AB}
\bibinfo{author}{{NXP Semiconductors}}, \bibinfo{year}{2018}.
\newblock \bibinfo{title}{{Datasheet: i.MX 6Solo CPU (MCIMX6S5EVM10AB)}}.
\newblock \URLprefix
  \url{https://www.nxp.com/docs/en/data-sheet/IMX6SDLCEC.pdf}.
  \bibinfo{note}{{Accessed: 2023-10-09}}.
\bibitem[{{Robert Bosch GmbH}(2023a)}]{bosch_connect}
\bibinfo{author}{{Robert Bosch GmbH}}, \bibinfo{year}{2023}a.
\newblock \bibinfo{title}{{Bosch eBike Connect}}.
\newblock \URLprefix \url{https://www.ebike-connect.com/}.
  \bibinfo{note}{{Accessed: 2023-10-01}}.
\bibitem[{{Robert Bosch GmbH}(2023b)}]{bosch_oss}
\bibinfo{author}{{Robert Bosch GmbH}}, \bibinfo{year}{2023}b.
\newblock \bibinfo{title}{{Bosch eBike Systems Licences Products}}.
\newblock \URLprefix \url{https://www.bosch-ebike.com/de/licences-products}.
  \bibinfo{note}{{Accessed: 2023-10-01}}.
\bibitem[{{Robert Bosch GmbH}(2023c)}]{bosch_tool}
\bibinfo{author}{{Robert Bosch GmbH}}, \bibinfo{year}{2023}c.
\newblock \bibinfo{title}{{Index of /data/DiagnosisSoftware/Update}}.
\newblock \URLprefix
  \url{http://bosch-ebike-updates.com/data/DiagnosisSoftware/Update/}.
  \bibinfo{note}{accessed: 2023-10-01}.
\bibitem[{Strandberg et~al.(2023)Strandberg, Nowdehi and
  Olovsson}]{DBLP:journals/tiv/StrandbergNO23}
\bibinfo{author}{Strandberg, K.}, \bibinfo{author}{Nowdehi, N.},
  \bibinfo{author}{Olovsson, T.}, \bibinfo{year}{2023}.
\newblock \bibinfo{title}{A systematic literature review on automotive digital
  forensics: Challenges, technical solutions and data collection}.
\newblock \bibinfo{journal}{{IEEE} Trans. Intell. Veh.} \bibinfo{volume}{8},
  \bibinfo{pages}{1350--1367}.
\newblock \URLprefix \url{https://doi.org/10.1109/TIV.2022.3188340},
  \DOIprefix\doi{10.1109/TIV.2022.3188340}.
\bibitem[{{TheRoundup.org}(2023)}]{ebike_stats}
\bibinfo{author}{{TheRoundup.org}}, \bibinfo{year}{2023}.
\newblock \bibinfo{title}{{51 Official Ebike Statistics \& Facts}}.
\newblock \URLprefix \url{https://theroundup.org/ebike-statistics/}.
  \bibinfo{note}{{Accessed: 2023-10-10}}.
\bibitem[{Villarreal et~al.(2022)Villarreal, Verma, Upton and
  Beebe}]{villarreal2022nondestructive}
\bibinfo{author}{Villarreal, A.M.}, \bibinfo{author}{Verma, R.K.},
  \bibinfo{author}{Upton, O.}, \bibinfo{author}{Beebe, N.L.},
  \bibinfo{year}{2022}.
\newblock \bibinfo{title}{Nondestructive data acquisition methodology for iot
  devices: A case study on amazon echo dot version 2}.
\newblock \bibinfo{journal}{IEEE Internet of Things Journal}
  \bibinfo{volume}{10}, \bibinfo{pages}{4375--4387}.
\bibitem[{{Yocto Project}(2023)}]{yocto:poky}
\bibinfo{author}{{Yocto Project}}, \bibinfo{year}{2023}.
\newblock \bibinfo{title}{{Poky}}.
\newblock \URLprefix \url{https://www.yoctoproject.org/software-item/poky/}.
  \bibinfo{note}{{Accessed: 2023-10-09}}.
\bibitem[{Youn et~al.(2021)Youn, Lim, Seo, Chung and Lee}]{youn2021forensic}
\bibinfo{author}{Youn, M.A.}, \bibinfo{author}{Lim, Y.}, \bibinfo{author}{Seo,
  K.}, \bibinfo{author}{Chung, H.}, \bibinfo{author}{Lee, S.},
  \bibinfo{year}{2021}.
\newblock \bibinfo{title}{Forensic analysis for ai speaker with display echo
  show 2nd generation as a case study}, in: \bibinfo{booktitle}{Proceedings of
  the Digital Forensics Research Conference (DFRWS APAC)}.

\end{thebibliography}

\printcredits

\end{document}